\documentclass[prmaterials,superscriptaddress, floatfix, letterpaper,twocolumn,aps,showpacs]{revtex4-2}
\UseRawInputEncoding
\usepackage{inputenc}
\usepackage{natbib}
\usepackage{amsmath}
\usepackage{amsbsy}
\usepackage{amssymb}
\usepackage{scrextend}
\usepackage{graphicx}
\usepackage{color}
\usepackage{xcolor}
\usepackage{float}
\usepackage{siunitx} 
\usepackage[colorlinks=true, allcolors = blue]{hyperref}
\usepackage{soul}
\usepackage{tabularx}
\usepackage{array}
\newcolumntype{L}[1]{>{\raggedright\let\newline\\\arraybackslash\hspace{0pt}}m{#1}}
\newcolumntype{C}[1]{>{\centering\let\newline\\\arraybackslash\hspace{0pt}}m{#1}}
\newcolumntype{R}[1]{>{\raggedleft\let\newline\\\arraybackslash\hspace{0pt}}m{#1}}
\newcommand{\m}[1]{\begin{pmatrix} #1 \end{pmatrix}}

\newcommand{\BNA}{{$\mathrm{Ba}\mathrm{Ni}_{2}\mathrm{As}_{2}$}}

\newcommand{\CFA}{{$\mathrm{Ca}\mathrm{Fe}_{2}\mathrm{As}_{2}$}}
\newcommand{\CNA}{{$\mathrm{Ca}\mathrm{Ni}_{2}\mathrm{As}_{2}$}}
\newcommand{\SNA}{{$\mathrm{Sr}\mathrm{Ni}_{2}\mathrm{As}_{2}$}}
\newcommand{\TCS}{{$\mathrm{Th}\mathrm{Cr}_{2}\mathrm{Si}_{2}$}}
\newcommand{\BCNA}{{$\mathrm{Ba}_{1-x}\mathrm{Ca}_{x}\mathrm{Ni}_{2}\mathrm{As}_{2}$}}
\newcommand{\BSNA}{{$\mathrm{Ba}_{1-x}\mathrm{Sr}_{x}\mathrm{Ni}_{2}\mathrm{As}_{2}$}}

\newcommand{\BNAP}{{$\mathrm{Ba}\mathrm{Ni}_{2} \mathrm{(\mathrm{As}_{1-x}\mathrm{P}_{x})}_{2}$}}

\newcommand{\celsius}{{}^{\circ}\mathrm{C}}
\newcommand{\degreeCelsius}{\celsius}

\begin{document}
\title{Chemical pressure tuning of competing orders in $\textrm{Ba}_{1-x}\textrm{Ca}_{x}\textrm{Ni}_{2}\textrm{As}_{2}$}
\author{F. Henssler}
\affiliation{Institute for Quantum Materials and Technologies, Karlsruhe Institute of Technology, Kaiserstr. 12, D-76131 Karlsruhe, Germany}
\author{K.~Willa}
\affiliation{Institute for Quantum Materials and Technologies, Karlsruhe Institute of Technology, Kaiserstr. 12, D-76131 Karlsruhe, Germany}
\author{M.~Frachet}
\affiliation{Institute for Quantum Materials and Technologies, Karlsruhe Institute of Technology, Kaiserstr. 12, D-76131 Karlsruhe, Germany}
\affiliation{CEA-Leti, Universit\'{e} Grenoble Alpes, 17 avenue de Martyrs,
F-38054 Grenoble, France}
\author{T.~Lacmann}
\affiliation{Institute for Quantum Materials and Technologies, Karlsruhe Institute of Technology, Kaiserstr. 12, D-76131 Karlsruhe, Germany}
\author{D.~A.~Chaney}
\affiliation{ESRF, The European Synchrotron, 71 avenue des Martyrs, CS 40220 F-38043 Grenoble Cedex 9, France}
\author{M.~Merz}
\affiliation{Institute for Quantum Materials and Technologies, Karlsruhe Institute of Technology, Kaiserstr. 12, D-76131 Karlsruhe, Germany}
\affiliation{Karlsruhe Nano Micro Facility (KNMFi), Karlsruhe Institute of Technology, Kaiserstr. 12, D-76131 Karlsruhe, Germany}
\author{A.-A. Haghighirad}
\email{amir-abbas.haghighirad@kit.edu}
\affiliation{Institute for Quantum Materials and Technologies, Karlsruhe Institute of Technology, Kaiserstr. 12, D-76131 Karlsruhe, Germany}
\author{M. Le Tacon}
\email{matthieu.letacon@kit.edu}
\affiliation{Institute for Quantum Materials and Technologies, Karlsruhe Institute of Technology, Kaiserstr. 12, D-76131 Karlsruhe, Germany}

\date{\today}

\begin{abstract}
{\BNA}, a structural-analogue to the iron-based parent compound BaFe$_2$As$_2$, offers a unique platform to study the interplay between superconductivity, charge density waves and, possibly, electronic nematicity. Here, we report on the growth and characterization of {\BCNA} single crystals with $0 \leq x \leq 0.1$, using a combination of x-ray diffraction, diffuse x-ray scattering, heat capacity, and electronic transport measurements. Our results demonstrate that calcium substitution affects the structural, electronic and thermodynamic properties of {\BNA} in a way that is strongly reminiscent of moderate hydrostatic pressures albeit with marked differences. In particular Ca-substitution efficiently suppresses both the triclinic structural transition and the associated commensurate charge density wave formation, while increasing the superconducting transition temperature. We found that the substitution range in which the crystals remain homogeneous is limited as for concentrations $x \geq 0.04$ intense diffuse x-ray scattering indicates the formation of stacking faults, which, despite the preserved integrity of the NiAs layers, prevents investigation up to concentrations at which the chemical pressure would completely suppress the structural instability. 

\end{abstract}

\maketitle

\section{Introduction}

The coexistence of magnetically or charge-ordered phases with superconductivity is a common feature in the complex phase diagrams of quantum materials. A major challenge in this field is to unravel the nature of the interplay between these electronic phases~\cite{Fernandes19}. This is crucial for understanding unconventional superconductivity, where the emergent properties of quantum materials often depend on a delicate balance and interplay of different electronic orders. This balance can be experimentally tuned using controllable parameters such as pressure, strain, magnetic field, or charge carrier concentration~\cite{Basov17}.
Chemical substitution is one of the most widely used and effective methods for exploring the electronic landscape of unconventional superconducting materials, including high-temperature superconducting cuprates~\cite{Keimer_Nature2015} and iron-based superconductors~\cite{Stewart2011}. The present work focuses on the superconducting compound {\BNA}~\cite{Ronning08}, which at room temperature shares the {\TCS} tetragonal \emph{I}4\emph{/mmm} crystal structure with BaFe$_2$As$_2$, the parent compound of iron-based superconductors. However, {\BNA} exhibits a very different electronic phase diagram. Recently, it has attracted significant attention for hosting multiple charge density waves (as opposed to magnetic phases, which have not been detected so far) combined with structural distortions~\cite{Lee19, Lee21, Merz21, Meingast22, Yao22, Lacmann23, Souliou22, Song23}. The interplay between these features and the superconducting phase remains elusive.
Earlier studies of {\BNA} revealed a strong first-order structural transition to a triclinic ($P\bar{1}$) phase below $T_\mathrm{S} = \SI{135}{\mathrm{K}}$ (upon cooling), followed by a low-temperature superconducting instability at $T_{c} = \SI{0.6}{\mathrm{K}}$, presumably of conventional BCS type~\cite{Kurita09}. More recently, it was discovered that a long-range commensurate charge density wave (C-CDW) with a characteristic wavevector of $q = ($1/3 0 -1/3$)_\mathrm{tet}$ (for simplicity, all reciprocal space wavevector are given in the tetragonal notation $(H$ $K$ $L)_\mathrm{tet}$, but the subscript "$\mathrm{tet}$" will be omitted) forms in the triclinic phase~\cite{Lee19}. Additionally, the existence of an incommensurate and uniaxial charge density wave (I-CDW) at $q = ($0.28 0 0$)$ above the triclinic transition was reported and investigated in detail \cite{Lee19,Merz21}. Precursor diffuse x-ray scattering linked to this I-CDW has been observed even at room temperature and is associated with soft-phonon modes that condense upon cooling~\cite{Souliou22, Song23} and yield a long-range I-CDW order and a small $B_{1g}$-symmetric orthorhombic (\emph{Immm}) distortion below $\sim \SI{142}{\mathrm{K}}$ \citep{Meingast22, Merz21}.
The electronic phase diagram of {\BNA} has been explored using hydrostatic pressure \cite{Lacmann23, Collini23} and various isoelectronic chemical substitutions~\cite{Kudo12, Meingast22, Eckberg20, Frachet22, Lee21, Merz21}. Interestingly, irrespective of the crystallographic site on which these substitutions occur, $T_{S}$  decreases and, as the C-CDW gets eventually completely suppressed, a sudden increase of the superconducting transition temperature $T_c$ to $\sim$ 3.5 K~\cite{Kudo12, Meingast22, Eckberg20} occurs. The origin of this behavior is still debated. 
On the one hand, a recent study on the Sr-substituted compound {\BSNA} revealed the presence of a novel C-CDW with $q = ($0 1/2 1/2$)$\cite{Lee21}, while electrical resistivity measurements under strain were interpreted as evidence of electronic nematicity in the normal state of this compound and suggested that the enhancement of $T_c$ could occur through nematic fluctuations \cite{Eckberg20,Lee21}. 
On the other hand, elastoresistivity and thermodynamic investigations in phosphorus substituted samples revealed an overall different (but equally rich) phase diagram \cite{Frachet22,Meingast22,Yao22}, in which unusual nematic fluctuations are reported but also point to an absence of nematic criticality. Instead, these studies associate the increase of $T_c$ to an enhancement of the electron-phonon coupling in a more conventional electron-phonon pairing mechanism \cite{Subedi08,Ronning08,Kudo12,Song24}. 

The available data indicate that the electronic phase of {\BNA} is particularly sensitive to the lattice parameter ratio $c/a$~\citep{Meingast22}, and the different behavior may be related to the fact that this ratio is affected very differently by the different types of substitutions.
Indeed, crystallographic data show that substituting $\mathrm{Ba}$ with $\mathrm{Sr}$ in {\BSNA} yields a monotonic decrease of the $c/a$ ratio from 2.81 in pristine {\BNA} (at room temperature) to 2.45 in SrNi$_2$As$_2$, effectively resulting in a strong uniaxial $c$-axis chemical pressure. On the contrary, $c/a$ remains essentially constant in the case of substitution of $\mathrm{As}$ with $\mathrm{P}$~\cite{Meingast22}.
Furthermore, it is worth noting that whereas only $\sim \SI{7}{\%}$ of P substitution for As is needed to completely suppress the triclinic phase~\cite{Eckberg20, Kudo12, Noda17, Meingast22}, no less than $\sim\SI{70}{\%}$ of Sr on the Ba site are required to obtain the same effect. These disparities highlight that beyond chemical pressure, substitution could also significantly impact the electronic properties of {\BNA} through, e.g., the introduction of disorder, calling in turn for careful studies of the evolution of lattice and electronic degrees of freedom of {\BNA} under chemical substitution.

In this work, we explore the alternative isovalent substitution of Ba with Ca rather than with Sr. In the 122 Fe-based family, the electronic properties of {\CFA} exhibit an unusually strong pressure dependence. A moderate hydrostatic pressure of $\sim \SI{0.35}{\mathrm{GPa}}$ can turn the low temperature antiferromagnetic orthorhombic phase into a non-magnetically ordered collapsed tetragonal structure with a dramatic decrease of the unit cell volume by $\SI{5}{\%}$ and of the $c/a$ ratio by $\SI{11}{\%}$ ~\cite{Kreyssig08}.  
In the {\CNA} end member of the nickel series, the $c/a$ ratio of the tetragonal unit cell is identical to that of {\SNA}, albeit with a significant reduction of the total unit cell volume (from $\SI{177.6}{\AA^3}$ in {\SNA} to $\SI{164}{\AA^3}$ in {\CNA}~\cite{Mewis80}). This also suggests that a complete suppression of the triclinic/C-CDW phases could be achieved with lower substitution levels using Ca instead of Sr. 
\newline
We report on the crystal structure and characterization of the basic electronic properties of {\BCNA} single crystals up to $x \sim 0.1$. We obtained high quality, homogeneous single-crystals up to a concentration of $x\sim0.04$ and characterized them using energy-dispersive x-ray spectroscopy (EDX), x-ray diffraction (XRD), diffuse x-ray scattering (DS), electrical transport and heat capacity measurements. At higher Ca concentrations, we find a strong compression of the $a$-axis parameter, however, the crystals tend to be less homogeneous and reciprocal space reconstructions reveal long streaks along the reciprocal $L$ directions, a characteristic signature of stacking faults. We find that $\SI{9}{\%}$ of $\mathrm{Ca}$ substitution suppresses the triclinic phase transition temperature down to about \SI{100}{\mathrm{K}} (which would require about $\SI{50}{\%}$ of $\mathrm{Sr}$) and increases the superconducting transition temperature by $\sim \SI{50}{\%}$. In the investigated Ca-concentration range, the substitution yields a compression of lattice parameters similar to that obtained under hydrostatic pressure up to $\sim \SI{1.5}{\mathrm{GPa}}$~\cite{Lacmann23}. It also has a similar impact on the incommensurability of the I-CDW ($q_\textrm{I-CDW}$ slightly increases). In contrast with hydrostatic pressure, no additional CDW phase nor structural distortion were observed, as summarized in a first phase diagram of {\BCNA}. 

\section{Experimental Methods}
\label{sec:methods}
\begin{figure}[t]{
    \centering
    \includegraphics[width=0.482\textwidth]{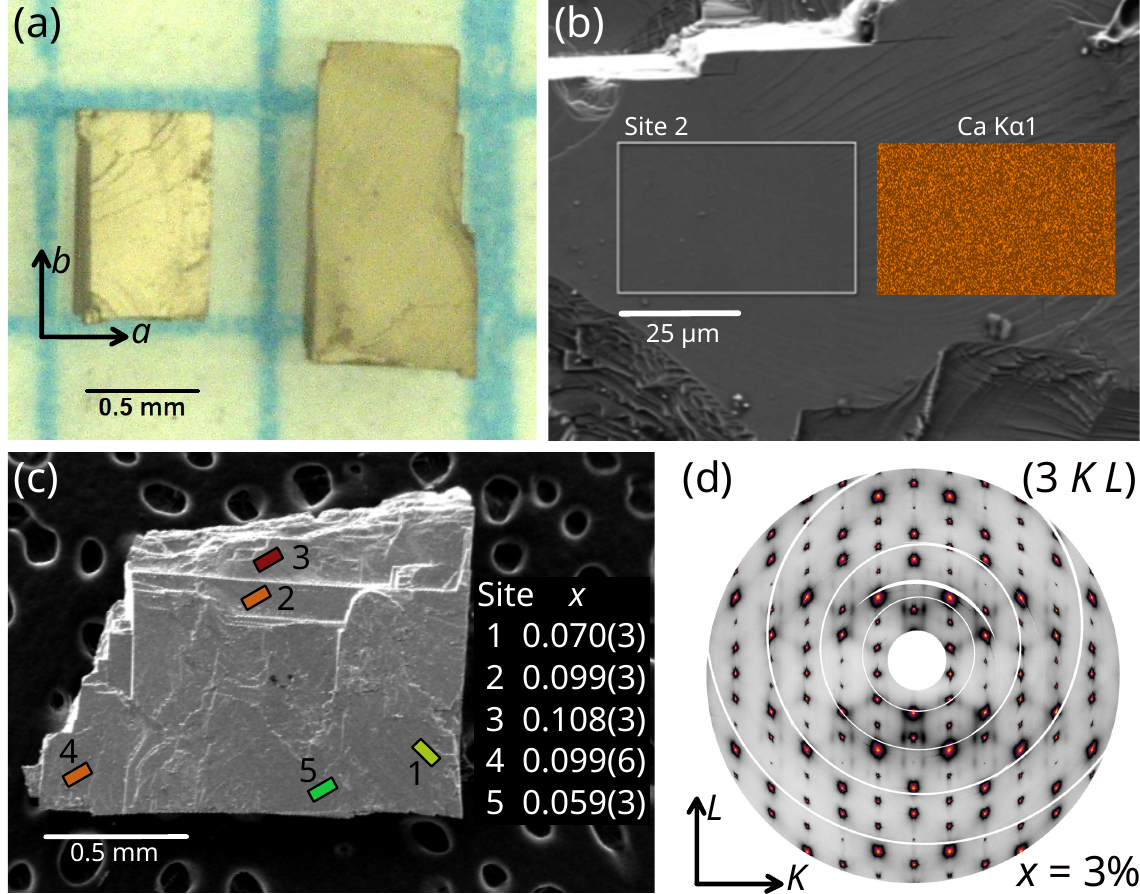} 
    \caption{(a) Typical {\BCNA} samples (here $x\sim0.02$). The sample sizes range up to $2 \times 2 \times \SI{0.3}{\mm^3}$. They are well-formed in a platelet shape, ideally with rectangular edges and dark surfaces shining with a gold luster. (b) SEM image of a {\BCNA} sample, measured at a voltage of $\SI{15}{kV}$ and a current of $\SI{20}{nA}$. On the right side, the intensity distribution of the $\mathrm{Ca}$ K$\alpha$1 line is depicted. (c) SEM image of a highly substituted {\BCNA} sample. Each area is between  $50 \times \SI{25}{\mu \m^2}$  to $100 \times \SI{50}{\mu \m^2}$ large (see Site 2 in (b)). The color indicates substitution level from green ($x\sim 0.06$) to red ($x\sim 0.11$). (d) Reconstruction of the reciprocal space plane $($3 $K$ $L)$ of an $x=0.03$ {\BCNA} sample, measured at $\SI{300}{K}$ with diffuse x-ray scattering at the ID28 beamline (ESRF).}
    \label{fig: crystals}
    }
\end{figure}

\begin{table}[b]
    \centering
    \caption{Growth parameters varied between the different batches.}
    \begin{tabular}{ccccccccc} 
    \hline
$x_\mathrm{Ca,nom}$ & $T_\mathrm{max} (\unit{\celsius}$) & $t_\mathrm{dwell}$(h) & $r_\mathrm{cool} (\unit{\celsius / \hour})$ \\
    \hline
0.05 & 1150 & 8 & 0.80\\
0.1 & 1180 & 8 & 0.77\\
0.2 & 1235 & 6 & 0.83\\
0.3 & 1235 & 6 & 0.80\\
0.4 & 1225 & 6 & 0.65\\
    \hline
    \end{tabular}
    \label{tab:Growth table}
\end{table}
The {\BCNA} single crystals were grown using a self-flux ($\mathrm{NiAs}$) method. All the handling operations were carried out in a glove box under inert Ar atmosphere (O$_\mathrm{2}< \SI{0.5}{ppm}$). The precursor $\mathrm{NiAs}$ was synthesized by mixing and grinding elemental $\mathrm{Ni}$ powder (Alfa Aesar $\SI{99.9}{\%}$) and $\mathrm{As}$ powder (chemPUR $\SI{99.9999}{\%}$). The mixture was then sealed in an evacuated quartz tube, which was subsequently heated to $\SI{730}{\degreeCelsius}$ with $\SI{50}{\degreeCelsius / h}$. After a dwell time of $\SI{20}{h}$ it was cooled down to $\SI{75}{\degreeCelsius}$ with $\SI{200}{\degreeCelsius / h}$.
The resulting $\mathrm{NiAs}$ was then ground to powder and mixed with $\mathrm{Ba}$ (lump, Alfa Aesar $\SI{99.99}{\%}$) and $\mathrm{Ca}$ (lump, Alfa Aesar $\SI{99.98}{\%}$) in the ratio $(\mathrm{Ba},\mathrm{Ca})$:$\mathrm{NiAs}$=1:4. Then, the powder was heated in a glassy carbon or alumina crucible placed in a sealed fused silica ampule (in vacuum) with $\SI{100}{\degreeCelsius / h}$ to $\SI{500}{\degreeCelsius}$ (dwell time $\SI{10}{h}$). The temperature was then increased with $\SI{100}{\degreeCelsius / h}$ to $T_\mathrm{max} = 1150\text{-}\SI{1235}{\degreeCelsius}$ for a dwell time of $6\text{-}\SI{8}{h}$ before it was slowly cooled (cooling rate $0. 65\text{-}\SI{0.83}{\degreeCelsius / h}$), followed by a decanting of the furnace at $\SI{980}{\degreeCelsius}$ (see Table \ref{tab:Growth table}, in which the growth parameters that were varied between the different batches are listed). A picture of a representative sample ($x\sim0.02$) is shown in Fig. \ref{fig: crystals}(a). 
\\
The chemical compositions of the obtained mm-sized single crystals (see Fig.~\ref{fig: crystals}(a)) were examined with a COXEM EM-30AXN benchtop device for scanning electron microscopy (SEM) and EDX (Fig. \ref{fig: crystals}(b) and (c)).
XRD experiments were conducted using a STOE imaging plate diffraction system (IPDS-2T) utilizing Mo $K_\mathrm{\alpha}$ radiation. 
For the investigated specimens all accessible reflections ($\approx 4600$)\@ were measured up to a maximum angle 2$\theta$ = 65$^{\circ}$ at a detector distance of 80 mm.\@ The data were corrected for Lorentz, polarization, extinction, and absorption effects by employing the STOE X-AREA software package. Complementing data sets were collected on our in-house high-flux, high-resolution, rotating anode RIGAKU Synergy-DW (Mo/Ag) diffractometer with Mo $K_\mathrm{\alpha}$ radiation.\@ The system is equipped with a background-less Hypix-Arc150$^{\circ}$ detector, which guarantees minimal reflection profile distortion and ensures uniform detection conditions for all reflections. For these measurements around 8400 reflections were collected in a detector distance of 47 mm and with a resolution up to 0.37 \AA,\@ data reduction was performed employing the CrysAlisPro software package~\cite{CrysalysPro}.\@ Using SHELXL \cite{Sheldrick08} and JANA2006 \cite{Petříček14} all available averaged symmetry-independent reflections (I $> 2\sigma$) have been included for the respective RT refinements in space group $I4/mmm$.\@ The refinements converged quite well and show very good weighted reliability factors ($wR_2$) typically between 4 and 8 \%,\@ depending on the mosaic spread and the stacking faults present in the individual samples. 
Additionally, DS measurements were performed at the ID28 beamline of the European Synchrotron Radiation Facility (ESRF)~\cite{Girard19}. The incident photon energy was $\SI{17.797}{keV}$, beam size was 40 $\times \SI{40}{\mu m^2}$ and the data was recorded using a Pilatus3 X 1M detector in shutterless mode with an angular step of 0.25$^{\circ}$. The sample detector distance was $\SI{244}{mm}$ and data was collected at two detector positions $\SI{19}{^\circ}$ and $\SI{48}{^\circ}$. Initial analysis was performed using the CrysAlis software package~\cite{CrysalysPro} and in-house software developed at the ID28 beamline was employed for high-quality reconstructions of selected reciprocal space layers.

Electrical transport measurements were performed on a Quantum Design Physical Property Measurement System (PPMS) or with a custom setup in an Oxford He-flow cryostat using a Keithley 6221 current source and a Keithley 2182A nanovoltmeter in delta mode. All investigated samples were electrically contacted in a standard four-probe geometry using Pt wires and silver paint (either DuPont 4929N or Hans Wolbring Leitsilber) or silver epoxy (EPO-TEK H20E). Resistance was measured in the $ab$-plane with a fixed current of $\SI{1}{mA}$ and a typical cooling rate of $\SI{1}{K/min}$. At low temperature the applied current was reduced to $\SI{0.2}{mA}$.

Specific heat $C_p$ measurements were performed using a PPMS with He-3 insert. Generally, large heat pulses (temperature rise $\sim \SI{30}{\%}$ of the absolute temperature) were evaluated with the dual-slope method except for the region around first-order transitions, where the single slope method was used. In order to achieve good thermal coupling, the samples were mounted on edge with a little amount of Apiezon N grease.

\section{Results}
\subsection{Crystal structure and composition}
The $\mathrm{Ca}$ concentration of the  {\BCNA} single crystals was determined from EDX measurements and from refinement of the XRD data. 
We observed that samples containing less than $\SI{4}{\%}$ of $\mathrm{Ca}$ ($x\leq 0.04$) were typically homogeneous and that the $\mathrm{Ca}$ concentration from sample to sample was found to be larely constant within the batch. Room temperature in-house XRD and synchrotron DS measurements confirmed the high crystalline quality of these single crystals, illustrated by the diffuse map of the $($3 $K$ $L)$ reciprocal plane shown in Fig. \ref{fig: crystals}(d), in which sharp Bragg reflections were observed in all three directions of the reciprocal space. 

\begin{figure}{
    \centering
    \includegraphics[width=0.482\textwidth]{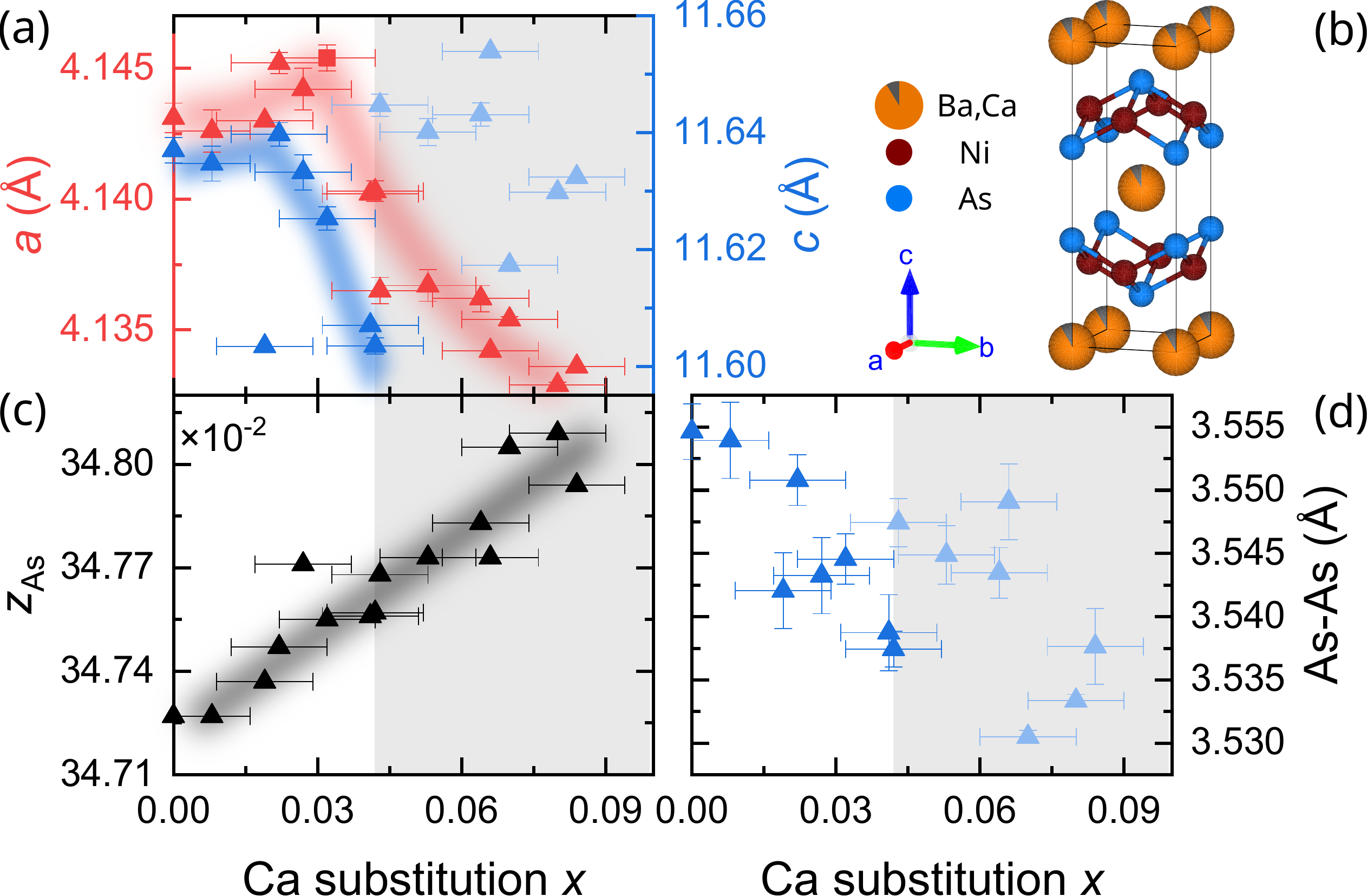} 
   \caption{(a) Substitution dependence of the lattice parameters \emph{a} and \emph{c} at room temperature. Samples with $\mathrm{Ca}$ concentration $x>0.04$ (shaded area) were found to have stacking faults that prevented accurate determination of the \emph{c}-axis lattice parameter (see light blue points). (b) The tetragonal {\BCNA} unit cell. (c) The dimensionless $z$ parameter of the Wyckoff position 4e, occupied by $\mathrm{As}$, as a function of the $\mathrm{Ca}$ concentration. (d) Substitution dependence of the out-of-plane As-As distance at room temperature. Error bars shown in the plots are statistical errors from the refinements.}
    \label{fig: lattice parameters}
    }
\end{figure}

The evolution of the lattice parameters as function of $\mathrm{Ca}$ concentration as well as of a selection of internal structural degrees of freedom, especially the height of $\mathrm{As}$, $z_{As}$, as the most relevant one, and the As-As distance along $c$, given by $(1 - 2 \times z_{As}) \times c$ 
are reported in Fig. \ref{fig: lattice parameters}. They show a relatively limited impact of Ca on the structure up to $x \sim 0.025$, where a slight increase of $a$ and $c$ parameters can be seen. Above this concentration, a sizeable contraction of these lattice parameters occurs. In parallel, we observe a continuous increase of $z_{As}$ with the $\mathrm{Ca}$ concentration, alongside a trend towards a small decrease of the interlayer As-As distance. 

\begin{figure}{
    \centering
    \includegraphics[width=0.45\textwidth]{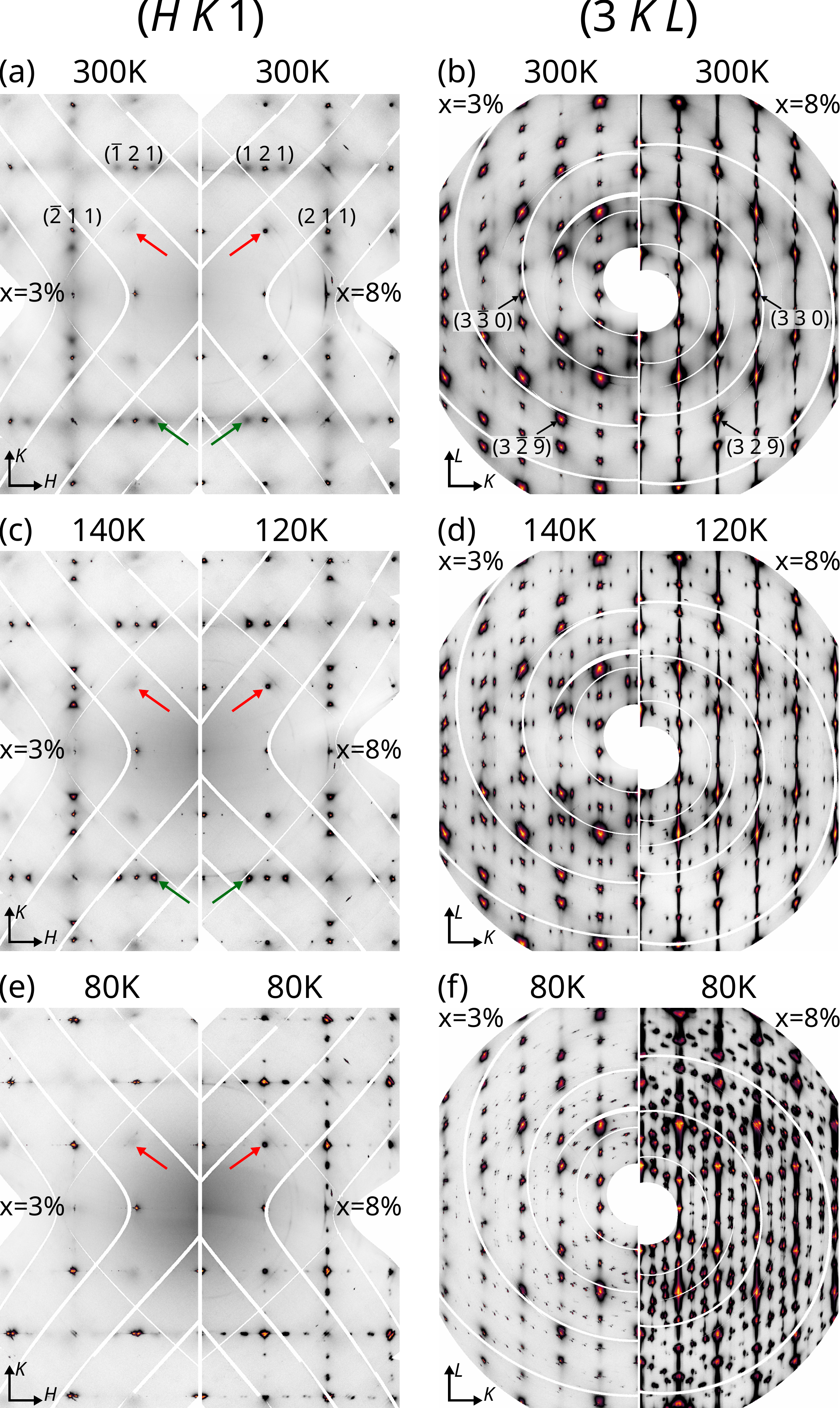} 
   \caption{Reciprocal space planes measured with diffuse x-ray scattering at the ID28 beamline (ESRF). The $(H$~$K$~1$)$ (a,c,e) and $($3 $K$ $L)$ (b,d,f) maps are shown for $x=0.03$ and $x=0.08$ for three characteristic temperatures. At room temperature {\BCNA} shows weak diffuse scattering at $q=$~(0.28~0~0$)$. On cooling, these I-CDW satellite reflections increase dramatically in intensity at about $\SI{140}{K}$ and $\SI{120}{K}$ for $x=0.03,0.08$, respectively (see green arrows in (a),(c)). At the base temperature, $\SI{80}{K}$, both samples are triclinic (reciprocal maps are shown in the tetragonal unit cell for comparison) and exhibit the C-CDW. In contrast to the lower doped sample, the $0.08$ sample shows clear streaks in the $($3 $K$ $L)$ maps, indicating stacking faults/disorder along the crystallographic \emph{c}-direction. The $(H$ $K$ 1$)$ maps, which are perpendicular to the $($3 $K$ $L)$, naturally show these streaks as a weaker additional peak at a symmetry forbidden position (see red arrows in (a), (c), (e)).}
    \label{fig: DS data}
    }
\end{figure}

The evaluation of other internal parameters, such as the Ni-As distance or the Ni-As-Ni bond angles, show very little dependence on Ca concentration, indicating little, if any, modification of the internal structure of the NiAs layers.
Importantly, with increasing $\mathrm{Ca}$ concentration, the determination of the $c$-axis parameter becomes less reliable. Indeed, as illustrated in Fig. \ref{fig: DS data}, where various diffuse scattering reciprocal space cuts for the $x=0.03$ and 0.08 samples are shown, long streaks along the \emph{L}-direction appear in the out-of-plane reciprocal space cuts above $x>0.04$. This indicates the presence of stacking faults in the structure for $x>0.04$. These also give rise to finite intensities at forbidden Bragg reflections in cuts of reciprocal layers perpendicular to the streaks (e.g. some intensity is seen at the (1 1 1) position - marked as red arrows in Fig. \ref{fig: DS data}(a)). Overall this results in a general degradation of the crystal periodic structure along the $c$-axis, which limits the accuracy of the $c$ lattice parameter estimate and of the internal structural parameters. It is in particular not possible to discuss in detail the role of the $c/a$ ratio, which has been reported in the {\BNAP} systems as an essential parameter controlling the electronic phase of this family of compounds~\cite{Meingast22}. Note that the reflections remain well defined in the H and K directions throughout the entire series.

We have not detected signatures of parasitic phases or inclusions in the investigated samples, but we note that for those with higher $x$ values, stronger variations of the $\mathrm{Ca}$ concentration from sample to sample within a given batch are reported with larger crystals pieces tending to be less homogeneous. EDX elemental maps on these samples did not show particular inhomogeneities in the $\mathrm{Ca}$ concentration over several tens of microns  on a given terrace, but 1-$\SI{2}{\%}$ variations of this concentration from one terrace to the next were routinely encountered (Fig.~\ref{fig: crystals} (c)).

\subsection{Incommensurate and Commensurate CDWs}
\begin{figure}{
    \centering
    \includegraphics[width=0.482\textwidth]{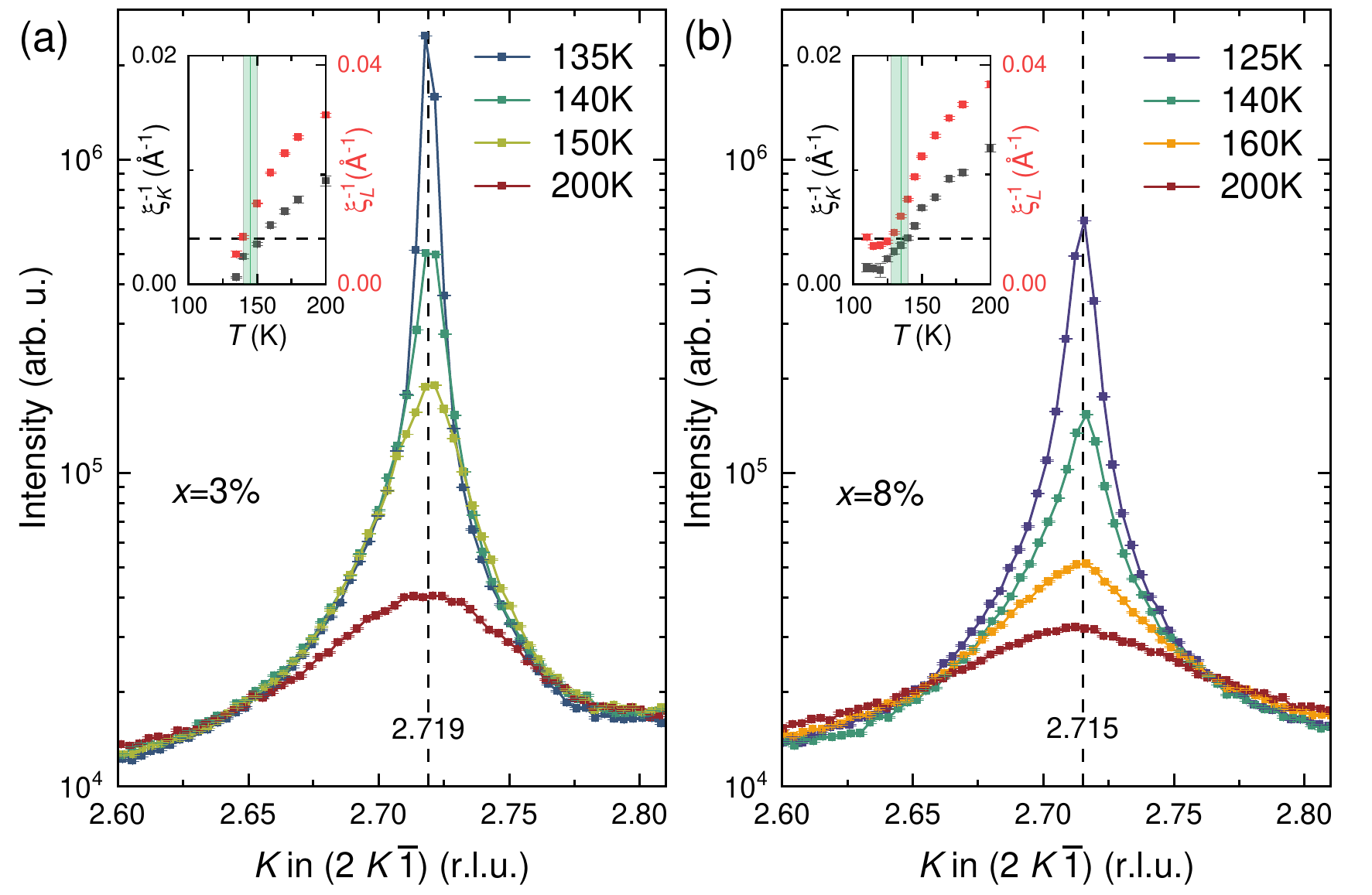} 
   \caption{Linecuts of the diffuse signal across $q_\textrm{I-CDW}$ along the [0$K$0] direction at serveral temperatures. In (a) the I-CDW peak is shown at 200, 150, 140 and $\SI{135}{K}$ for an $x \sim 0.03$ sample. The vertical dashed line gives the $K$ coordinate of the I-CDW peak center. The inset panel visualizes the inverse correlation length and the transition temperature $T_\mathrm{ICDW}$ in green, while the dashed line indicates the correlation length $\xi_\mathrm{K,L} = 250,\SI{120}{\mathrm{\AA}}$. In (b) the I-CDW peak of the $x \sim 0.08$ sample is presented the same style.}
    \label{fig: ICDW}
    }
\end{figure}

As previously reported for the parent compounds~\cite{Souliou22}, a strong diffuse signal can already be seen at room temperature in the Ca-rich samples at reciprocal space wave vectors close to  $q_{\textrm{I-CDW}}\sim($0.28~0~0$)$, at which the I-CDW develops at lower temperatures.
For all the investigated $\mathrm{Ca}$ concentrations, these develop into a long-range charge order, evidenced by the appearance of sharp Bragg-like reflections on top of the diffuse background upon cooling (Fig.~\ref{fig: ICDW}). Using this criterion, we can estimate that the formation of the long-range I-CDW takes place at around $\SI{145}{\mathrm{K}}$ for $x = 0.03$ and $\SI{135}{\mathrm{K}}$ for $x = 0.08$. As shown in the insets of Figure \ref{fig: ICDW}, in both compounds this occurs when the in-plane (resp. out-of-plane) correlation length reaches about $\xi_\mathrm{K} \sim \SI{250} {\mathrm{\AA}}$ ($\xi_\mathrm{L} \sim \SI{120} {\mathrm{\AA}}$), comparable with previous report on the parent compound~\cite{Souliou22}.
We further note a small change of the incommensurability, where the ordering vector increases from  $q=0.281$ ($x=0.03$) to $q=0.285$ ($x=0.08$).
At the lowest temperature, at which the x-ray measurements were carried out ($\SI{80}{K}$), the I-CDW in all samples investigated in this study has evolved into a C-CDW with an ordering vector $q_{\textrm{C-CDW}}=($1/3 0 -1/3$)$, which is --- as for the undoped compound \cite{Merz21} --- associated with the triclinic $P\bar{1}$ phase. No other superstructures were observed down to $\SI{80}{\mathrm{K}}$ for ($x \leq 0.084$).

\subsection{Phase transition temperatures}
\label{sec: electrical transport}

\begin{figure}{
    \centering
    \includegraphics[width=0.483\textwidth]{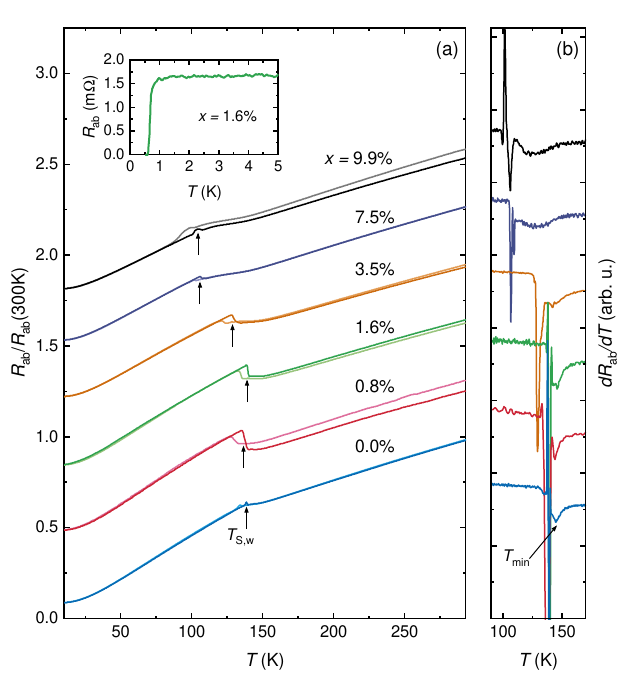} 
    \caption{(a) Temperature dependence of the in-plane electrical resistance of {\BCNA}, normalized by the room temperature value. The arrows indicate $T_\mathrm{S}$ on warming. The inset shows the resistance drop at the superconducting transition for {\BCNA} with $x=0.016$. (b) Temperature dependence of $dR_{ab}/dT$. The temperature of the local minimum $T_\mathrm{min}$ is indicated by an arrow for {\BNA}. The curves are offset for clarity.}
    \label{fig: resistance}
    }
\end{figure}
\begin{table}
    \centering
    \caption{Residual resistance ratio $RRR$ for several Ca substitutions $x$.}
    \begin{center}
    \begin{tabularx}{240pt}{>{\centering\arraybackslash}X|>{\centering\arraybackslash}X|>{\centering\arraybackslash}X|>{\centering\arraybackslash}X|>{\centering\arraybackslash}X|>{\centering\arraybackslash}X|>{\centering\arraybackslash}X|>{\centering\arraybackslash}X|>{\centering\arraybackslash}X} 
    \hline
$x(\%)$   & 0 & 0.8  & 1.3  & 1.6 & 3.5  & 7.5  & 9.4  & 9.9 \\
\hline
$RRR$  & 11.8 & 5.9 & 5.2 & 4.9 & 3.8 & 4.0 & 4.4 & 4.4 \\
    \hline
    \end{tabularx}
    \end{center}
    \label{tab:RRR}
\end{table}
Next, we turn to electrical transport measurements, which are shown in Fig.~\ref{fig: resistance}(a). These reveal a metallic behavior at all $\mathrm{Ca}$ concentrations. 
For our pure {\BNA} single crystals, the residual resistance ratio $RRR = R_{ab}(\SI{300}{\mathrm{K}})/R_{ab}(T \rightarrow 0))$ is close to 12, indicating high crystal quality in comparison to crystals previously grown by self-flux ($RRR=8$ \cite{Sefat09},\cite{Kudo12} and $RRR=5.5$  \cite{Eckberg20}) or $\mathrm{Pb}$-flux ($RRR=5$ \cite{Ronning08}). As the substitution level increases, $RRR$ rapidly decreases to $RRR=6$ at $x_{\rm{Ca}} \sim 0.01$ (see Table \ref{tab:RRR}), before it stabilizes at about $RRR=4$ up to the highest investigated concentrations. This is qualitatively in line with the comparably small  in-plane disorder inferred from x-ray measurements discussed above.
Indeed, in all the investigated samples a discontinuity in the resistance corresponding to first-order transition to the triclinic phase (concomitant with the appearance of the C-CDW discussed above) can be seen at $T=T_\mathrm{S}(x)$ \citep{Kudo12, Frachet22}. This feature is highlighted even more upon evaluating the derivative of the resistivity $dR/dT$, see Fig.~\ref{fig: resistance}(b). At this transition, the lattice parameters change abruptly yielding the formation of cracks in the sample or at the electrical contacts, which may offset the resistance (see discrepancy between cooling and warming curves in Fig ~\ref{fig: resistance}(a)).
More interesting is the $\mathrm{Ca}$ dependence of the transition temperature $T_S(x)$. Following the trend observed in the lattice parameters, $T_S$ displays only small changes up to $x\sim0.02$, before starting to rapidly decrease.
As previously noted ~\cite{Yao22} a local minimum in $dR_{ab}/dT$ is seen at $T_{min}$, a few K above $T_S$ at a temperature comparable to that, at which the long range I-CDW forms.
Our experimental set-up for electrical transport did not allow us to explore the superconducting transition systematically, but we could detect it in a $x=0.016$ crystal at $T_{c} \sim \SI{0.7}{\mathrm{K}}$, a temperature slightly above that of the pristine material. 

\label{sec: specific heat}
\begin{figure}[t]{
    \centering
    \includegraphics[width=0.482\textwidth]{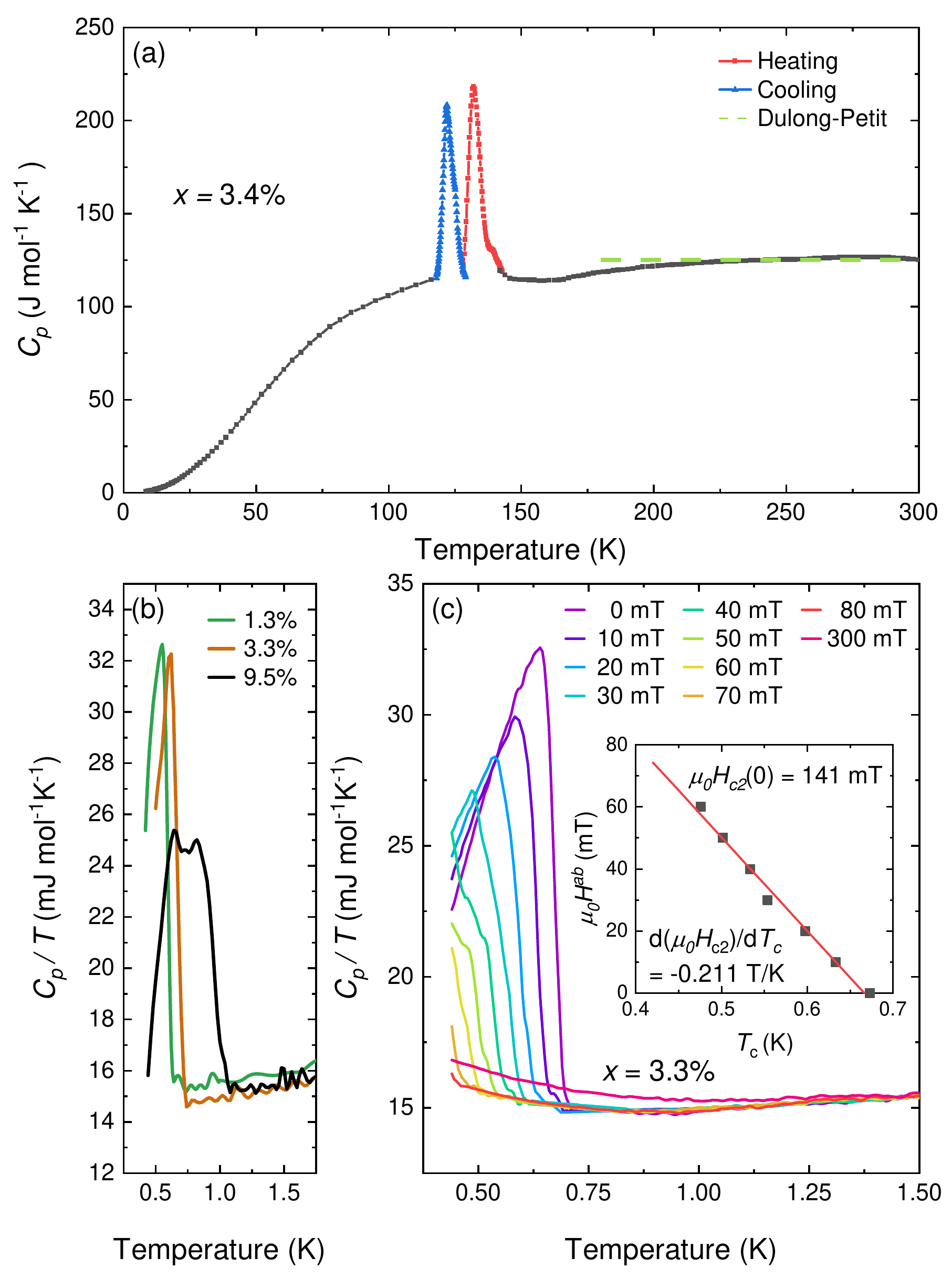} 
    \caption{(a) Temperature dependence of the specific heat of {\BCNA} for a sample with $x=0.034$. The blue line indicates the first-order transition peak at $T_\mathrm{S,c} \sim \SI{124}{\mathrm{K}}$ on cooling, the red one at $T_\mathrm{S,w} \sim \SI{135}{\mathrm{K}}$ on warming. At high temperature, $C_p$ reaches the Dulong-Petit limit $\sim 3p_\mathrm{uc}k_\mathrm{B}$ (with $k_\mathrm{B}$ being the Boltzmann constant and $p_\mathrm{uc}$ the number of atoms per unit cell), that is depicted by the green dashed line. (b) $C_p/T$ at low temperature. The discontinuity indicates bulk superconductivity below $\SI{0.6}{\mathrm{K}}$, $\SI{0.7}{\mathrm{K}}$ and $\SI{1}{\mathrm{K}}$ for $x=0.013, 0.033$ and $0.095$, respectively. (c) $C_p/T$ vs. T with varying in-plane magnetic field for a {\BCNA} sample with $x=0.033$. The inset shows the extracted $T_c$ as a function of the applied magnetic field, fitted linearly by the red line.}
    \label{fig: Specific Heat}
    }
\end{figure}
To gain additional insights on the $\mathrm{Ca}$ concentration dependence of $T_c$ and of the structural phase transitions in {\BCNA}, we have looked for their thermodynamic signatures by performing specific heat measurements in the temperature range from $\SI{300}{K}$ to $\SI{0.4}{K}$. Figure~\ref{fig: Specific Heat}(a) shows the specific heat $C_p$ for a {\BCNA} sample with $x=0.034$ from room temperature down to $\SI{2}{\mathrm{K}}$. At high temperature the $C_p$  saturates at the Dulong-Petit limit (see green dashed line) \cite{DulongPetit1819}. 
The first-order triclinic transition manifests as a pronounced peak \citep{Yao22, Meingast22, Kudo12}, showing a hysteretic behavior at temperatures that are in excellent agreement with $T_{S}$ inferred from transport and x-ray measurements discussed in the previous sections. Note that the limited size of the crystals available did not allow us to resolve the specific heat jump associated with the second-order transition of the tetragonal to the (minute) orthorhombic distortion evidenced in P-substituted compounds~\cite{Meingast22} as the long range I-CDW forms~\cite{Souliou22}.

Additionally, we have extracted the Sommerfeld coefficient $\gamma$ and the Debye temperature $\Theta_\mathrm{D}$ by fitting the low temperature range from $\SI{2}{\mathrm{K}}$ to $\SI{8}{\mathrm{K}}$ to a second-order polynomial $C_p(T)/T = \gamma + \beta T^2 + \alpha T^4 $. 

\subsection{Superconducting properties}
At low temperatures, a second-order phase transition marks the onset of superconductivity. The superconducting transition temperature $T_{c}$ is determined by applying the entropy conserving construction \cite{Willa18}. In the Fig.~\ref{fig: Specific Heat}(b) $C_p(T)/T $ is shown below $\SI{1.75}{\mathrm{K}}$ for three {\BCNA} samples with different $\mathrm{Ca}$ content. The specific heat jump of Ca substituted samples occurs at a slightly higher temperature than in the parent compound, confirming $T_{c}$ obtained from the low temperature resistivity (see inset Fig. \ref{fig: resistance}(a)).
While the increase of $T_{c}$ with $x$ is weak but systematic, $\Delta C/ (\gamma T_{c})\sim 1.4$ remains essentially constant in the substitution range $x\leq 0.033$. This is in good agreement with the undoped compound as well as with standard weak-coupling BCS theory \cite{BCS57}.
For high substitution levels, we observe that the onset of the superconducting transition temperature is significantly increased, while the transition broadens substantially compared to lower substitution levels. It could in fact be interpreted as two sharp superconducting transitions with respective $T_{c}$ of about $\SI{0.7}{\mathrm{K}}$ and $\SI{1.0}{\mathrm{K}}$ (Considering only the later onset naturally yields a decrease of the $\Delta C/(\gamma T_{c})$ ratio). This is again consistent with inhomogeneous $\mathrm{Ca}$ distribution for $x>0.04$ samples.

The $T_{c}, \gamma, \Theta_\mathrm{D}$ and $\Delta C/ (\gamma T_{c})$ extracted from our analysis are summarized together with the samples respective $\mathrm{Ca}$ content and mass in the Tab.~\ref{tab: HC overview}. The Debye temperature increases slowly with $\mathrm{Ca}$ concentration, in contrast with the observations reported in the cases of Sr \cite{Eckberg20}, Cu \cite{Kudo17}, and P \cite{Kudo12} substitutions, where the $\Theta_\mathrm{D}$ remains essentially constant before reducing notably following the suppression of the triclinic phase.
Conversely, in the case of Co \cite{Eckberg18} substitutions, $\Theta_\mathrm{D}$ exhibits minimal change across the entire doping series. 
The Sommerfeld coefficient $\gamma$ on the other hand appears essentially independent of the  $\mathrm{Ca}$ concentration, akin to observation made in the cases of Cu and P substitutions, but in contrast with Co and Sr substitutions, which respectively slightly increase and decrease $\gamma$.
 
\setlength{\tabcolsep}{3.1pt}
\def\arraystretch{1.1}
\begin{table}[t]
    \centering
    \caption{Superconducting transition temperature $T_{c}$, Sommerfeld coefficient $\gamma$, Debye-temperature $\Theta_\mathrm{D}$ and the ratio $\Delta C/(\gamma T_{c})$ of several {\BCNA} samples with mass $m$. The first row shows results from \cite{Sefat09}, the second from \cite{Kudo12}. The last row shows results of a sample in the stacking faults regime. }
    \begin{tabular}{cccccc} 
    \hline
$x$ &  $m\,\mathrm{(mg)}$ & $T_{c}\,\mathrm{(K)}$ & $\gamma\,\mathrm{(mJ/mol\, K^2)}$ & $\Theta_\mathrm{D}\,\mathrm{(K)}$ & $\Delta C/\gamma\, T_{c}$ \\
    \hline

0.0 & & 0.7 & 13.2 & 250 & 1.38\\
0.0 & & 0.6 & 14.0 & 250 & 1.30\\
0.013 & 3.45 & 0.6 & 14.8 & 275 & 1.36 \\
0.033 & 2.77 & 0.7 & 14.4 & 280 & 1.40 \\
0.095 & 1.07 & 1.0 & 14.3 & 260 & 0.78 \\
\hline
    \end{tabular}
    \label{tab: HC overview}
\end{table}

Finally, we discuss the field dependence of the superconducting transition temperature by looking at the low temperature specific heat of a $x=0.033$ sample with an in-plane applied magnetic field (see Fig. \ref{fig: Specific Heat}(c)). The resulting $T_{c}( \mu_0 H)$ is plotted in the inset. From a linear fit we obtain an upper critical field slope of $d(\mu_0H_{c2})/dT_{c}=\SI{-0.211}{\mathrm{T/K}}$, which is similar to that of the pristine {\BNA} ($\SI{-0.396}{\mathrm{T/K}}$ \cite{Ronning08}, $\SI{-0.226}{\mathrm{T/K}}$ \cite{Kurita09}), but much smaller than what is observed in iron-based superconductors, where $d(\mu_0H_{c2})/dT_{c}\sim\SI{-5}{\mathrm{T/K}}$ \cite{Hardy20, Willa19, Prozorov24}. We can then estimate $H_{c2}(0)$ with $H(T_{c})= 0.7 \cdot \left( H_{c2}(0)-dH_{c2}/dT_{c} \cdot T_{c} \right)$ \cite{Werthamer66}, which results in an upper critical field of $\mu_0 H_\mathrm{c2}(0) = \SI{141}{\mathrm{mT}}$ again similar to {\BNA} \cite{Ronning08}.

\section{Discussion}

\begin{figure}[b]{
    \centering
    \includegraphics[width=0.45\textwidth]{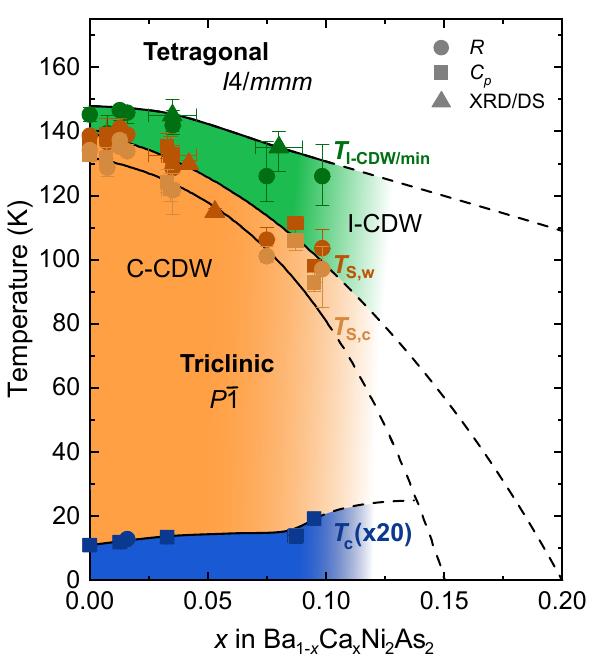} 
    \caption{Phase diagram of {\BCNA} up to a substitution level of about $x=0.1$. Circles represent results from electrical transport, squares from specific heat, and triangles from XRD and DS measurements. }
    \label{fig: Phase diagram}
    }
\end{figure}

We start our discussion by summarizing our experimental results in the phase diagram shown in Fig.~\ref{fig: Phase diagram}. After a small initial increase, $T_\mathrm{S,c}$ (measured on cooling, hereafter referred to as $T_\mathrm{S}$) decreases continuously from about $\SI{137}{\mathrm{K}}$ to below $\SI{100}{\mathrm{K}}$. In the entire substitution range investigated in this work, the phase transition into the triclinic phase remains hysteretic and first order. Above it, we detect an I-CDW phase similar to that of the parent compound, with a small increase of the incommensureate ordering vector $q_\textrm{I-CDW}$ with $x$. 
At low temperature, our thermodynamic and transport measurements detect a superconducting transition at a temperature that increases with $\mathrm{Ca}$ content from $\SI{0.6}{\mathrm{K}}$ to $\SI{1}{\mathrm{K}}$. No additional structural or electronic phases were detected in the investigated substitution range.

The decrease of $T_S$ with increasing Ca concentration we report is qualitatively similar to the effects observed upon P-substitution on the As site, or Sr-substitution on the Ba site \citep{Kudo12, Meingast22,Eckberg20}. Focusing on the Ba-site substitutions, we note that quantitatively Ca substitutions are significantly more effective in suppressing $T_S$ than Sr. Indeed $T_\mathrm{S}=\SI{120}{\mathrm{K}}$ is reached with only $x_\mathrm{Ca}\sim 0.035$, while $\mathrm{Sr}$-concentration of about $x_\mathrm{Sr}\sim 0.3$ are needed. Similarly, we reached $T_\mathrm{S}=\SI{100}{\mathrm{K}}$ with $x_\mathrm{Ca}\sim 0.09$, which would take $x_\mathrm{Sr}\sim 0.5$\cite{Eckberg20}. A simple extrapolation suggests that a concentration between 15 and $\SI{20}{\%}$ of Ca would be sufficient to suppress completely the triclinic/C-CDW phase.
Given that these substitutions are isovalent, this difference cannot be attributed to charge-doping effects. Furthermore, since the most efficient suppression is achieved in the system requiring the least amount of substitution, we can conclude that the rate of $T_S$ decrease is not controlled by the associated cationic disorder.
In comparison to the Sr case, which mostly impacts the $c$-axis, while leaving $a$ barely affected up to $x_\mathrm{Sr}\sim 0.7$~\cite{Eckberg20}, a contraction of both $a$- and $c$-axis lattice parameters can be seen already from $x_\mathrm{Ca}\sim 0.03$. This suggests that the modifications induced by Ca substitutions in {\BNA} are most likely rooted in chemical pressure effects.
Along this line, we note that Ca-doping of the Ba site appears unique since it preserves the NiAs layers, and both, $c$ and $a$, (globally) decrease, mimicking the effect of hydrostatic pressure. A direct comparison of the relative lattice parameters changes induced by pressure~\cite{Lacmann23} indicate that $\SI{3}{\%}$ of $\mathrm{Ca}$ have the same effect on the sample unit cell as $\sim \SI{0.5}{\mathrm{GPa}}$ of hydrostatic pressure.
The rather weak dependence of the onset temperature of the I-CDW transition and the increase of the ordering vector $q_\textrm{I-CDW}$ from 0.28 to 0.285 at $x=0.08$ are indeed comparable to the effect of a hydrostatic compression of $\sim \SI{1.5}{\mathrm{GPa}}$~\cite{Lacmann23}. 
Nonetheless, the suppression of $T_S$ and the small increase of the superconducting $T_c$ upon Ca-substitution are stronger than the reported effects of pressure \cite{Park_2010} and reminiscent of the Sr and P-substituted cases. 
Importantly, in this same regime Sr- and P- substituted samples show a large $B_{1g}$ elastoresistance signal, whose possible electronic nematic origin is debated \cite{Frachet22, Eckberg20, Meingast22}. Although no such experiments have been performed yet on Ca-doped samples, the strong similarity of freestanding electrical resistance of Ca- and P-substituted samples strongly suggests a similar behavior. Our specific heat results (Fig. \ref{fig: Specific Heat}(c)) further indicate that within this doping range superconductivity remains in the weak-coupling BCS limit. Finally, the limited doping range investigated did not allow us to fully suppress the coincident C-CDW and triclinic phase transition, where $T_c$ has been reported to sharply increase to $\sim 3$K in P-~\cite{Kudo12, Meingast22} or Sr-~\cite{Eckberg20} substituted samples. An optimization of the growth conditions appears needed to reach these optimal conditions for superconductivity while avoiding stacking faults.
This might be improved by increasing the growth temperature above 1250$^{\circ}$C, which however calls for the use of alternative growth conditions (crucible material, gas pressure, etc.). Furthermore, the increased vapour pressure of binary arsenides, CaAs and NiAs used in the growth at these elevated temperatures probably requires high-pressure synthesis similar to that used for BaNi$_2$P$_2$~\cite{Kito08, Tomioka08}. 

\section{Summary}
In conclusion, we report the growth and a detailed investigation of \BCNA~single crystals, up to $x \sim 0.1$. Upon Ca-substitution, akin to earlier studies on Sr- and P-substituted samples, the triclinic and associated C-CDW transition are suppressed while the superconducting one is enhanced. In contrast, and especially compared to the effect of Sr-substitution, which also occurs on the Ba site, the effect of Ca on the crystal structure appears to be much closer to that of hydrostatic pressure~\cite{Lacmann23}. As a consequence, the amount of Ca required to suppress the triclinic transition temperature is significantly lower than that of Sr, which opens a promising avenue to study the possible nematicity of the compound at lower disorder level. 
Our thermodynamic measurements strongly suggest that, within the entire range investigated, superconductivity remains of weak-coupling BCS type. However, for $x_{\textrm{Ca}} \geq 0.04$ we observe the evidence of the formation of numerous stacking faults, which limits the range of our investigation and in particular of the relevance of the $c/a$ ratio in the phase tuning. While this observation calls for the optimization of the growth conditions, we stress that this should not preclude investigating the possible electronic nematicity of {\BCNA}.
\newline

\begin{acknowledgments}
Fruitful discussions with F. Hardy, C. Meingast and S.-M. Souliou are gratefully acknowledged. We acknowledge support by the Deutsche Forschungsgemeinschaft (DFG; German Research Foundation) under CRC/TRR 288 (Project B03). We acknowledge the European Synchrotron Radiation Facility (ESRF) for provision of synchrotron radiation facilities under proposal number HC-4946. K.W. acknowledges support from the Swiss National Science Foundation through the Postdoc.Mobility program. M.F. acknowledges funding from the Alexander von Humboldt Foundation and the Young Investigator Group preparatory program of the Karlsruhe Institute for Technology. 
\end{acknowledgments}

\bibliographystyle{apsrev4-2}

\bibliography{BCNA.bib}

\begin{thebibliography}{40}%
\makeatletter
\providecommand \@ifxundefined [1]{%
 \@ifx{#1\undefined}
}%
\providecommand \@ifnum [1]{%
 \ifnum #1\expandafter \@firstoftwo
 \else \expandafter \@secondoftwo
 \fi
}%
\providecommand \@ifx [1]{%
 \ifx #1\expandafter \@firstoftwo
 \else \expandafter \@secondoftwo
 \fi
}%
\providecommand \natexlab [1]{#1}%
\providecommand \enquote  [1]{``#1''}%
\providecommand \bibnamefont  [1]{#1}%
\providecommand \bibfnamefont [1]{#1}%
\providecommand \citenamefont [1]{#1}%
\providecommand \href@noop [0]{\@secondoftwo}%
\providecommand \href [0]{\begingroup \@sanitize@url \@href}%
\providecommand \@href[1]{\@@startlink{#1}\@@href}%
\providecommand \@@href[1]{\endgroup#1\@@endlink}%
\providecommand \@sanitize@url [0]{\catcode `\\12\catcode `\$12\catcode `\&12\catcode `\#12\catcode `\^12\catcode `\_12\catcode `\%12\relax}%
\providecommand \@@startlink[1]{}%
\providecommand \@@endlink[0]{}%
\providecommand \url  [0]{\begingroup\@sanitize@url \@url }%
\providecommand \@url [1]{\endgroup\@href {#1}{\urlprefix }}%
\providecommand \urlprefix  [0]{URL }%
\providecommand \Eprint [0]{\href }%
\providecommand \doibase [0]{https://doi.org/}%
\providecommand \selectlanguage [0]{\@gobble}%
\providecommand \bibinfo  [0]{\@secondoftwo}%
\providecommand \bibfield  [0]{\@secondoftwo}%
\providecommand \translation [1]{[#1]}%
\providecommand \BibitemOpen [0]{}%
\providecommand \bibitemStop [0]{}%
\providecommand \bibitemNoStop [0]{.\EOS\space}%
\providecommand \EOS [0]{\spacefactor3000\relax}%
\providecommand \BibitemShut  [1]{\csname bibitem#1\endcsname}%
\let\auto@bib@innerbib\@empty
\bibitem [{\citenamefont {Fernandes}\ \emph {et~al.}(2019)\citenamefont {Fernandes}, \citenamefont {Orth},\ and\ \citenamefont {Schmalian}}]{Fernandes19}%
  \BibitemOpen
  \bibfield  {author} {\bibinfo {author} {\bibfnamefont {R.~M.}\ \bibnamefont {Fernandes}}, \bibinfo {author} {\bibfnamefont {P.~P.}\ \bibnamefont {Orth}},\ and\ \bibinfo {author} {\bibfnamefont {J.}~\bibnamefont {Schmalian}},\ }\href {https://doi.org/https://doi.org/10.1146/annurev-conmatphys-031218-013200} {\bibfield  {journal} {\bibinfo  {journal} {Annual Review of Condensed Matter Physics}\ }\textbf {\bibinfo {volume} {10}},\ \bibinfo {pages} {133} (\bibinfo {year} {2019})}\BibitemShut {NoStop}%
\bibitem [{\citenamefont {Basov}\ \emph {et~al.}(2017)\citenamefont {Basov}, \citenamefont {Averitt},\ and\ \citenamefont {Hsieh}}]{Basov17}%
  \BibitemOpen
  \bibfield  {author} {\bibinfo {author} {\bibfnamefont {D.~N.}\ \bibnamefont {Basov}}, \bibinfo {author} {\bibfnamefont {R.~D.}\ \bibnamefont {Averitt}},\ and\ \bibinfo {author} {\bibfnamefont {D.}~\bibnamefont {Hsieh}},\ }\href {https://doi.org/10.1038/nmat5017} {\bibfield  {journal} {\bibinfo  {journal} {Nature Materials}\ }\textbf {\bibinfo {volume} {16}},\ \bibinfo {pages} {1077} (\bibinfo {year} {2017})}\BibitemShut {NoStop}%
\bibitem [{\citenamefont {Keimer}\ \emph {et~al.}(2015)\citenamefont {Keimer}, \citenamefont {Kivelson}, \citenamefont {Norman}, \citenamefont {Uchida},\ and\ \citenamefont {Zaanen}}]{Keimer_Nature2015}%
  \BibitemOpen
  \bibfield  {author} {\bibinfo {author} {\bibfnamefont {B.}~\bibnamefont {Keimer}}, \bibinfo {author} {\bibfnamefont {S.~A.}\ \bibnamefont {Kivelson}}, \bibinfo {author} {\bibfnamefont {M.~R.}\ \bibnamefont {Norman}}, \bibinfo {author} {\bibfnamefont {S.}~\bibnamefont {Uchida}},\ and\ \bibinfo {author} {\bibfnamefont {J.}~\bibnamefont {Zaanen}},\ }\href {https://doi.org/10.1038/nature14165} {\bibfield  {journal} {\bibinfo  {journal} {Nature}\ }\textbf {\bibinfo {volume} {518}},\ \bibinfo {pages} {179} (\bibinfo {year} {2015})}\BibitemShut {NoStop}%
\bibitem [{\citenamefont {Stewart}(2011)}]{Stewart2011}%
  \BibitemOpen
  \bibfield  {author} {\bibinfo {author} {\bibfnamefont {G.~R.}\ \bibnamefont {Stewart}},\ }\href {https://doi.org/10.1103/RevModPhys.83.1589} {\bibfield  {journal} {\bibinfo  {journal} {Rev. Mod. Phys.}\ }\textbf {\bibinfo {volume} {83}},\ \bibinfo {pages} {1589} (\bibinfo {year} {2011})}\BibitemShut {NoStop}%
\bibitem [{\citenamefont {Ronning}\ \emph {et~al.}(2008)\citenamefont {Ronning}, \citenamefont {Kurita}, \citenamefont {Bauer}, \citenamefont {Scott}, \citenamefont {Park}, \citenamefont {Klimczuk}, \citenamefont {Movshovich},\ and\ \citenamefont {Thompson}}]{Ronning08}%
  \BibitemOpen
  \bibfield  {author} {\bibinfo {author} {\bibfnamefont {F.}~\bibnamefont {Ronning}}, \bibinfo {author} {\bibfnamefont {N.}~\bibnamefont {Kurita}}, \bibinfo {author} {\bibfnamefont {E.~D.}\ \bibnamefont {Bauer}}, \bibinfo {author} {\bibfnamefont {B.~L.}\ \bibnamefont {Scott}}, \bibinfo {author} {\bibfnamefont {T.}~\bibnamefont {Park}}, \bibinfo {author} {\bibfnamefont {T.}~\bibnamefont {Klimczuk}}, \bibinfo {author} {\bibfnamefont {R.}~\bibnamefont {Movshovich}},\ and\ \bibinfo {author} {\bibfnamefont {J.~D.}\ \bibnamefont {Thompson}},\ }\href {https://doi.org/10.1088/0953-8984/20/34/342203} {\bibfield  {journal} {\bibinfo  {journal} {Journal of Physics: Condensed Matter}\ }\textbf {\bibinfo {volume} {20}},\ \bibinfo {pages} {342203} (\bibinfo {year} {2008})}\BibitemShut {NoStop}%
\bibitem [{\citenamefont {Lee}\ \emph {et~al.}(2019)\citenamefont {Lee}, \citenamefont {de~la Pe\~na}, \citenamefont {Sun}, \citenamefont {Mitrano}, \citenamefont {Fang}, \citenamefont {Jang}, \citenamefont {Lee}, \citenamefont {Eckberg}, \citenamefont {Campbell}, \citenamefont {Collini}, \citenamefont {Paglione}, \citenamefont {de~Groot},\ and\ \citenamefont {Abbamonte}}]{Lee19}%
  \BibitemOpen
  \bibfield  {author} {\bibinfo {author} {\bibfnamefont {S.}~\bibnamefont {Lee}}, \bibinfo {author} {\bibfnamefont {G.}~\bibnamefont {de~la Pe\~na}}, \bibinfo {author} {\bibfnamefont {S.~X.~L.}\ \bibnamefont {Sun}}, \bibinfo {author} {\bibfnamefont {M.}~\bibnamefont {Mitrano}}, \bibinfo {author} {\bibfnamefont {Y.}~\bibnamefont {Fang}}, \bibinfo {author} {\bibfnamefont {H.}~\bibnamefont {Jang}}, \bibinfo {author} {\bibfnamefont {J.-S.}\ \bibnamefont {Lee}}, \bibinfo {author} {\bibfnamefont {C.}~\bibnamefont {Eckberg}}, \bibinfo {author} {\bibfnamefont {D.}~\bibnamefont {Campbell}}, \bibinfo {author} {\bibfnamefont {J.}~\bibnamefont {Collini}}, \bibinfo {author} {\bibfnamefont {J.}~\bibnamefont {Paglione}}, \bibinfo {author} {\bibfnamefont {F.~M.~F.}\ \bibnamefont {de~Groot}},\ and\ \bibinfo {author} {\bibfnamefont {P.}~\bibnamefont {Abbamonte}},\ }\href {https://doi.org/10.1103/PhysRevLett.122.147601} {\bibfield  {journal} {\bibinfo  {journal} {Physical Review Letters}\ }\textbf {\bibinfo {volume} {122}},\
  \bibinfo {pages} {147601} (\bibinfo {year} {2019})}\BibitemShut {NoStop}%
\bibitem [{\citenamefont {Lee}\ \emph {et~al.}(2021)\citenamefont {Lee}, \citenamefont {Collini}, \citenamefont {Sun}, \citenamefont {Mitrano}, \citenamefont {Guo}, \citenamefont {Eckberg}, \citenamefont {Paglione}, \citenamefont {Fradkin},\ and\ \citenamefont {Abbamonte}}]{Lee21}%
  \BibitemOpen
  \bibfield  {author} {\bibinfo {author} {\bibfnamefont {S.}~\bibnamefont {Lee}}, \bibinfo {author} {\bibfnamefont {J.}~\bibnamefont {Collini}}, \bibinfo {author} {\bibfnamefont {S.~X.~L.}\ \bibnamefont {Sun}}, \bibinfo {author} {\bibfnamefont {M.}~\bibnamefont {Mitrano}}, \bibinfo {author} {\bibfnamefont {X.}~\bibnamefont {Guo}}, \bibinfo {author} {\bibfnamefont {C.}~\bibnamefont {Eckberg}}, \bibinfo {author} {\bibfnamefont {J.}~\bibnamefont {Paglione}}, \bibinfo {author} {\bibfnamefont {E.}~\bibnamefont {Fradkin}},\ and\ \bibinfo {author} {\bibfnamefont {P.}~\bibnamefont {Abbamonte}},\ }\href {https://doi.org/10.1103/PhysRevLett.127.027602} {\bibfield  {journal} {\bibinfo  {journal} {Physical Review Letters}\ }\textbf {\bibinfo {volume} {127}},\ \bibinfo {pages} {027602} (\bibinfo {year} {2021})}\BibitemShut {NoStop}%
\bibitem [{\citenamefont {Merz}\ \emph {et~al.}(2021)\citenamefont {Merz}, \citenamefont {Wang}, \citenamefont {Wolf}, \citenamefont {Nagel}, \citenamefont {Meingast},\ and\ \citenamefont {Schuppler}}]{Merz21}%
  \BibitemOpen
  \bibfield  {author} {\bibinfo {author} {\bibfnamefont {M.}~\bibnamefont {Merz}}, \bibinfo {author} {\bibfnamefont {L.}~\bibnamefont {Wang}}, \bibinfo {author} {\bibfnamefont {T.}~\bibnamefont {Wolf}}, \bibinfo {author} {\bibfnamefont {P.}~\bibnamefont {Nagel}}, \bibinfo {author} {\bibfnamefont {C.}~\bibnamefont {Meingast}},\ and\ \bibinfo {author} {\bibfnamefont {S.}~\bibnamefont {Schuppler}},\ }\href {https://doi.org/10.1103/PhysRevB.104.184509} {\bibfield  {journal} {\bibinfo  {journal} {Physical Review B}\ }\textbf {\bibinfo {volume} {104}},\ \bibinfo {pages} {184509} (\bibinfo {year} {2021})}\BibitemShut {NoStop}%
\bibitem [{\citenamefont {Meingast}\ \emph {et~al.}(2022)\citenamefont {Meingast}, \citenamefont {Shukla}, \citenamefont {Wang}, \citenamefont {Heid}, \citenamefont {Hardy}, \citenamefont {Frachet}, \citenamefont {Willa}, \citenamefont {Lacmann}, \citenamefont {Le~Tacon}, \citenamefont {Merz}, \citenamefont {Haghighirad},\ and\ \citenamefont {Wolf}}]{Meingast22}%
  \BibitemOpen
  \bibfield  {author} {\bibinfo {author} {\bibfnamefont {C.}~\bibnamefont {Meingast}}, \bibinfo {author} {\bibfnamefont {A.}~\bibnamefont {Shukla}}, \bibinfo {author} {\bibfnamefont {L.}~\bibnamefont {Wang}}, \bibinfo {author} {\bibfnamefont {R.}~\bibnamefont {Heid}}, \bibinfo {author} {\bibfnamefont {F.}~\bibnamefont {Hardy}}, \bibinfo {author} {\bibfnamefont {M.}~\bibnamefont {Frachet}}, \bibinfo {author} {\bibfnamefont {K.}~\bibnamefont {Willa}}, \bibinfo {author} {\bibfnamefont {T.}~\bibnamefont {Lacmann}}, \bibinfo {author} {\bibfnamefont {M.}~\bibnamefont {Le~Tacon}}, \bibinfo {author} {\bibfnamefont {M.}~\bibnamefont {Merz}}, \bibinfo {author} {\bibfnamefont {A.-A.}\ \bibnamefont {Haghighirad}},\ and\ \bibinfo {author} {\bibfnamefont {T.}~\bibnamefont {Wolf}},\ }\href {https://doi.org/10.1103/PhysRevB.106.144507} {\bibfield  {journal} {\bibinfo  {journal} {Physical Review B}\ }\textbf {\bibinfo {volume} {106}},\ \bibinfo {pages} {144507} (\bibinfo {year} {2022})}\BibitemShut {NoStop}%
\bibitem [{\citenamefont {Yao}\ \emph {et~al.}(2022)\citenamefont {Yao}, \citenamefont {Willa}, \citenamefont {Lacmann}, \citenamefont {Souliou}, \citenamefont {Frachet}, \citenamefont {Willa}, \citenamefont {Merz}, \citenamefont {Weber}, \citenamefont {Meingast}, \citenamefont {Heid}, \citenamefont {Haghighirad}, \citenamefont {Schmalian},\ and\ \citenamefont {Le~Tacon}}]{Yao22}%
  \BibitemOpen
  \bibfield  {author} {\bibinfo {author} {\bibfnamefont {Y.}~\bibnamefont {Yao}}, \bibinfo {author} {\bibfnamefont {R.}~\bibnamefont {Willa}}, \bibinfo {author} {\bibfnamefont {T.}~\bibnamefont {Lacmann}}, \bibinfo {author} {\bibfnamefont {S.-M.}\ \bibnamefont {Souliou}}, \bibinfo {author} {\bibfnamefont {M.}~\bibnamefont {Frachet}}, \bibinfo {author} {\bibfnamefont {K.}~\bibnamefont {Willa}}, \bibinfo {author} {\bibfnamefont {M.}~\bibnamefont {Merz}}, \bibinfo {author} {\bibfnamefont {F.}~\bibnamefont {Weber}}, \bibinfo {author} {\bibfnamefont {C.}~\bibnamefont {Meingast}}, \bibinfo {author} {\bibfnamefont {R.}~\bibnamefont {Heid}}, \bibinfo {author} {\bibfnamefont {A.-A.}\ \bibnamefont {Haghighirad}}, \bibinfo {author} {\bibfnamefont {J.}~\bibnamefont {Schmalian}},\ and\ \bibinfo {author} {\bibfnamefont {M.}~\bibnamefont {Le~Tacon}},\ }\href {https://doi.org/10.1038/s41467-022-32112-7} {\bibfield  {journal} {\bibinfo  {journal} {Nature Communications}\ }\textbf {\bibinfo {volume} {13}},\ \bibinfo {pages}
  {4535} (\bibinfo {year} {2022})}\BibitemShut {NoStop}%
\bibitem [{\citenamefont {Lacmann}\ \emph {et~al.}(2023)\citenamefont {Lacmann}, \citenamefont {Haghighirad}, \citenamefont {Souliou}, \citenamefont {Merz}, \citenamefont {Garbarino}, \citenamefont {Glazyrin}, \citenamefont {Heid},\ and\ \citenamefont {Le~Tacon}}]{Lacmann23}%
  \BibitemOpen
  \bibfield  {author} {\bibinfo {author} {\bibfnamefont {T.}~\bibnamefont {Lacmann}}, \bibinfo {author} {\bibfnamefont {A.-A.}\ \bibnamefont {Haghighirad}}, \bibinfo {author} {\bibfnamefont {S.-M.}\ \bibnamefont {Souliou}}, \bibinfo {author} {\bibfnamefont {M.}~\bibnamefont {Merz}}, \bibinfo {author} {\bibfnamefont {G.}~\bibnamefont {Garbarino}}, \bibinfo {author} {\bibfnamefont {K.}~\bibnamefont {Glazyrin}}, \bibinfo {author} {\bibfnamefont {R.}~\bibnamefont {Heid}},\ and\ \bibinfo {author} {\bibfnamefont {M.}~\bibnamefont {Le~Tacon}},\ }\href {https://doi.org/10.1103/PhysRevB.108.224115} {\bibfield  {journal} {\bibinfo  {journal} {Physical Review B}\ }\textbf {\bibinfo {volume} {108}},\ \bibinfo {pages} {224115} (\bibinfo {year} {2023})}\BibitemShut {NoStop}%
\bibitem [{\citenamefont {Souliou}\ \emph {et~al.}(2022)\citenamefont {Souliou}, \citenamefont {Lacmann}, \citenamefont {Heid}, \citenamefont {Meingast}, \citenamefont {Frachet}, \citenamefont {Paolasini}, \citenamefont {Haghighirad}, \citenamefont {Merz}, \citenamefont {Bosak},\ and\ \citenamefont {Le~Tacon}}]{Souliou22}%
  \BibitemOpen
  \bibfield  {author} {\bibinfo {author} {\bibfnamefont {S.~M.}\ \bibnamefont {Souliou}}, \bibinfo {author} {\bibfnamefont {T.}~\bibnamefont {Lacmann}}, \bibinfo {author} {\bibfnamefont {R.}~\bibnamefont {Heid}}, \bibinfo {author} {\bibfnamefont {C.}~\bibnamefont {Meingast}}, \bibinfo {author} {\bibfnamefont {M.}~\bibnamefont {Frachet}}, \bibinfo {author} {\bibfnamefont {L.}~\bibnamefont {Paolasini}}, \bibinfo {author} {\bibfnamefont {A.~A.}\ \bibnamefont {Haghighirad}}, \bibinfo {author} {\bibfnamefont {M.}~\bibnamefont {Merz}}, \bibinfo {author} {\bibfnamefont {A.}~\bibnamefont {Bosak}},\ and\ \bibinfo {author} {\bibfnamefont {M.}~\bibnamefont {Le~Tacon}},\ }\href {https://doi.org/10.1103/PhysRevLett.129.247602} {\bibfield  {journal} {\bibinfo  {journal} {Physical Review Letters}\ }\textbf {\bibinfo {volume} {129}},\ \bibinfo {pages} {247602} (\bibinfo {year} {2022})}\BibitemShut {NoStop}%
\bibitem [{\citenamefont {Song}\ \emph {et~al.}(2023)\citenamefont {Song}, \citenamefont {Wu}, \citenamefont {Chen}, \citenamefont {He}, \citenamefont {Uchiyama}, \citenamefont {Li}, \citenamefont {Cao}, \citenamefont {Guo}, \citenamefont {Cao},\ and\ \citenamefont {Birgeneau}}]{Song23}%
  \BibitemOpen
  \bibfield  {author} {\bibinfo {author} {\bibfnamefont {Y.}~\bibnamefont {Song}}, \bibinfo {author} {\bibfnamefont {S.}~\bibnamefont {Wu}}, \bibinfo {author} {\bibfnamefont {X.}~\bibnamefont {Chen}}, \bibinfo {author} {\bibfnamefont {Y.}~\bibnamefont {He}}, \bibinfo {author} {\bibfnamefont {H.}~\bibnamefont {Uchiyama}}, \bibinfo {author} {\bibfnamefont {B.}~\bibnamefont {Li}}, \bibinfo {author} {\bibfnamefont {S.}~\bibnamefont {Cao}}, \bibinfo {author} {\bibfnamefont {J.}~\bibnamefont {Guo}}, \bibinfo {author} {\bibfnamefont {G.}~\bibnamefont {Cao}},\ and\ \bibinfo {author} {\bibfnamefont {R.}~\bibnamefont {Birgeneau}},\ }\href {https://doi.org/10.1103/PhysRevB.107.L041113} {\bibfield  {journal} {\bibinfo  {journal} {Physical Review B}\ }\textbf {\bibinfo {volume} {107}},\ \bibinfo {pages} {L041113} (\bibinfo {year} {2023})}\BibitemShut {NoStop}%
\bibitem [{\citenamefont {Kurita}\ \emph {et~al.}(2009)\citenamefont {Kurita}, \citenamefont {Ronning}, \citenamefont {Tokiwa}, \citenamefont {Bauer}, \citenamefont {Subedi}, \citenamefont {Singh}, \citenamefont {Thompson},\ and\ \citenamefont {Movshovich}}]{Kurita09}%
  \BibitemOpen
  \bibfield  {author} {\bibinfo {author} {\bibfnamefont {N.}~\bibnamefont {Kurita}}, \bibinfo {author} {\bibfnamefont {F.}~\bibnamefont {Ronning}}, \bibinfo {author} {\bibfnamefont {Y.}~\bibnamefont {Tokiwa}}, \bibinfo {author} {\bibfnamefont {E.~D.}\ \bibnamefont {Bauer}}, \bibinfo {author} {\bibfnamefont {A.}~\bibnamefont {Subedi}}, \bibinfo {author} {\bibfnamefont {D.~J.}\ \bibnamefont {Singh}}, \bibinfo {author} {\bibfnamefont {J.~D.}\ \bibnamefont {Thompson}},\ and\ \bibinfo {author} {\bibfnamefont {R.}~\bibnamefont {Movshovich}},\ }\href {https://doi.org/10.1103/PhysRevLett.102.147004} {\bibfield  {journal} {\bibinfo  {journal} {Physical Review Letters}\ }\textbf {\bibinfo {volume} {102}},\ \bibinfo {pages} {147004} (\bibinfo {year} {2009})}\BibitemShut {NoStop}%
\bibitem [{\citenamefont {Collini}\ \emph {et~al.}(2023)\citenamefont {Collini}, \citenamefont {Campbell}, \citenamefont {Sneed}, \citenamefont {Saraf}, \citenamefont {Eckberg}, \citenamefont {Jeffries}, \citenamefont {Butch},\ and\ \citenamefont {Paglione}}]{Collini23}%
  \BibitemOpen
  \bibfield  {author} {\bibinfo {author} {\bibfnamefont {J.}~\bibnamefont {Collini}}, \bibinfo {author} {\bibfnamefont {D.~J.}\ \bibnamefont {Campbell}}, \bibinfo {author} {\bibfnamefont {D.}~\bibnamefont {Sneed}}, \bibinfo {author} {\bibfnamefont {P.}~\bibnamefont {Saraf}}, \bibinfo {author} {\bibfnamefont {C.}~\bibnamefont {Eckberg}}, \bibinfo {author} {\bibfnamefont {J.}~\bibnamefont {Jeffries}}, \bibinfo {author} {\bibfnamefont {N.}~\bibnamefont {Butch}},\ and\ \bibinfo {author} {\bibfnamefont {J.}~\bibnamefont {Paglione}},\ }\href {https://doi.org/10.1103/PhysRevB.108.205103} {\bibfield  {journal} {\bibinfo  {journal} {Physical Review B}\ }\textbf {\bibinfo {volume} {108}},\ \bibinfo {pages} {205103} (\bibinfo {year} {2023})}\BibitemShut {NoStop}%
\bibitem [{\citenamefont {Kudo}\ \emph {et~al.}(2012)\citenamefont {Kudo}, \citenamefont {Takasuga}, \citenamefont {Okamoto}, \citenamefont {Hiroi},\ and\ \citenamefont {Nohara}}]{Kudo12}%
  \BibitemOpen
  \bibfield  {author} {\bibinfo {author} {\bibfnamefont {K.}~\bibnamefont {Kudo}}, \bibinfo {author} {\bibfnamefont {M.}~\bibnamefont {Takasuga}}, \bibinfo {author} {\bibfnamefont {Y.}~\bibnamefont {Okamoto}}, \bibinfo {author} {\bibfnamefont {Z.}~\bibnamefont {Hiroi}},\ and\ \bibinfo {author} {\bibfnamefont {M.}~\bibnamefont {Nohara}},\ }\href {https://doi.org/10.1103/PhysRevLett.109.097002} {\bibfield  {journal} {\bibinfo  {journal} {Phys Rev Lett}\ }\textbf {\bibinfo {volume} {109}},\ \bibinfo {pages} {097002} (\bibinfo {year} {2012})}\BibitemShut {NoStop}%
\bibitem [{\citenamefont {Eckberg}\ \emph {et~al.}(2020)\citenamefont {Eckberg}, \citenamefont {Campbell}, \citenamefont {Metz}, \citenamefont {Collini}, \citenamefont {Hodovanets}, \citenamefont {Drye}, \citenamefont {Zavalij}, \citenamefont {Christensen}, \citenamefont {Fernandes}, \citenamefont {Lee}, \citenamefont {Abbamonte}, \citenamefont {Lynn},\ and\ \citenamefont {Paglione}}]{Eckberg20}%
  \BibitemOpen
  \bibfield  {author} {\bibinfo {author} {\bibfnamefont {C.}~\bibnamefont {Eckberg}}, \bibinfo {author} {\bibfnamefont {D.~J.}\ \bibnamefont {Campbell}}, \bibinfo {author} {\bibfnamefont {T.}~\bibnamefont {Metz}}, \bibinfo {author} {\bibfnamefont {J.}~\bibnamefont {Collini}}, \bibinfo {author} {\bibfnamefont {H.}~\bibnamefont {Hodovanets}}, \bibinfo {author} {\bibfnamefont {T.}~\bibnamefont {Drye}}, \bibinfo {author} {\bibfnamefont {P.}~\bibnamefont {Zavalij}}, \bibinfo {author} {\bibfnamefont {M.~H.}\ \bibnamefont {Christensen}}, \bibinfo {author} {\bibfnamefont {R.~M.}\ \bibnamefont {Fernandes}}, \bibinfo {author} {\bibfnamefont {S.}~\bibnamefont {Lee}}, \bibinfo {author} {\bibfnamefont {P.}~\bibnamefont {Abbamonte}}, \bibinfo {author} {\bibfnamefont {J.~W.}\ \bibnamefont {Lynn}},\ and\ \bibinfo {author} {\bibfnamefont {J.}~\bibnamefont {Paglione}},\ }\href {https://doi.org/10.1038/s41567-019-0736-9} {\bibfield  {journal} {\bibinfo  {journal} {Nature Physics}\ }\textbf {\bibinfo {volume} {16}},\ \bibinfo
  {pages} {346} (\bibinfo {year} {2020})}\BibitemShut {NoStop}%
\bibitem [{\citenamefont {Frachet}\ \emph {et~al.}(2022)\citenamefont {Frachet}, \citenamefont {Wiecki}, \citenamefont {Lacmann}, \citenamefont {Souliou}, \citenamefont {Willa}, \citenamefont {Meingast}, \citenamefont {Merz}, \citenamefont {Haghighirad}, \citenamefont {Le~Tacon},\ and\ \citenamefont {B\"{o}hmer}}]{Frachet22}%
  \BibitemOpen
  \bibfield  {author} {\bibinfo {author} {\bibfnamefont {M.}~\bibnamefont {Frachet}}, \bibinfo {author} {\bibfnamefont {P.}~\bibnamefont {Wiecki}}, \bibinfo {author} {\bibfnamefont {T.}~\bibnamefont {Lacmann}}, \bibinfo {author} {\bibfnamefont {S.~M.}\ \bibnamefont {Souliou}}, \bibinfo {author} {\bibfnamefont {K.}~\bibnamefont {Willa}}, \bibinfo {author} {\bibfnamefont {C.}~\bibnamefont {Meingast}}, \bibinfo {author} {\bibfnamefont {M.}~\bibnamefont {Merz}}, \bibinfo {author} {\bibfnamefont {A.~A.}\ \bibnamefont {Haghighirad}}, \bibinfo {author} {\bibfnamefont {M.}~\bibnamefont {Le~Tacon}},\ and\ \bibinfo {author} {\bibfnamefont {A.~E.}\ \bibnamefont {B\"{o}hmer}},\ }\href {https://doi.org/10.1038/s41535-022-00525-8} {\bibfield  {journal} {\bibinfo  {journal} {npj Quantum Materials}\ }\textbf {\bibinfo {volume} {7}},\ \bibinfo {pages} {115} (\bibinfo {year} {2022})}\BibitemShut {NoStop}%
\bibitem [{\citenamefont {Subedi}\ and\ \citenamefont {Singh}(2008)}]{Subedi08}%
  \BibitemOpen
  \bibfield  {author} {\bibinfo {author} {\bibfnamefont {A.}~\bibnamefont {Subedi}}\ and\ \bibinfo {author} {\bibfnamefont {D.~J.}\ \bibnamefont {Singh}},\ }\href {https://doi.org/10.1103/PhysRevB.78.132511} {\bibfield  {journal} {\bibinfo  {journal} {Physical Review B}\ }\textbf {\bibinfo {volume} {78}},\ \bibinfo {pages} {132511} (\bibinfo {year} {2008})}\BibitemShut {NoStop}%
\bibitem [{\citenamefont {Song}\ \emph {et~al.}(2024)\citenamefont {Song}, \citenamefont {Si}, \citenamefont {Fennell}, \citenamefont {Stuhr}, \citenamefont {Deng}, \citenamefont {Wang}, \citenamefont {Liu}, \citenamefont {Hao}, \citenamefont {Luo}, \citenamefont {Liu}, \citenamefont {Meng},\ and\ \citenamefont {Li}}]{Song24}%
  \BibitemOpen
  \bibfield  {author} {\bibinfo {author} {\bibfnamefont {L.}~\bibnamefont {Song}}, \bibinfo {author} {\bibfnamefont {J.}~\bibnamefont {Si}}, \bibinfo {author} {\bibfnamefont {T.}~\bibnamefont {Fennell}}, \bibinfo {author} {\bibfnamefont {U.}~\bibnamefont {Stuhr}}, \bibinfo {author} {\bibfnamefont {G.}~\bibnamefont {Deng}}, \bibinfo {author} {\bibfnamefont {J.}~\bibnamefont {Wang}}, \bibinfo {author} {\bibfnamefont {J.}~\bibnamefont {Liu}}, \bibinfo {author} {\bibfnamefont {L.}~\bibnamefont {Hao}}, \bibinfo {author} {\bibfnamefont {H.}~\bibnamefont {Luo}}, \bibinfo {author} {\bibfnamefont {M.}~\bibnamefont {Liu}}, \bibinfo {author} {\bibfnamefont {S.}~\bibnamefont {Meng}},\ and\ \bibinfo {author} {\bibfnamefont {S.}~\bibnamefont {Li}},\ }\href {https://doi.org/10.1103/PhysRevB.109.104518} {\bibfield  {journal} {\bibinfo  {journal} {Physical Review B}\ }\textbf {\bibinfo {volume} {109}},\ \bibinfo {pages} {104518} (\bibinfo {year} {2024})}\BibitemShut {NoStop}%
\bibitem [{\citenamefont {Noda}\ \emph {et~al.}(2017)\citenamefont {Noda}, \citenamefont {Kudo}, \citenamefont {Takasuga}, \citenamefont {Nohara}, \citenamefont {Sugimoto}, \citenamefont {Ootsuki}, \citenamefont {Kobayashi}, \citenamefont {Horiba}, \citenamefont {Ono}, \citenamefont {Kumigashira}, \citenamefont {Fujimori}, \citenamefont {Saini},\ and\ \citenamefont {Mizokawa}}]{Noda17}%
  \BibitemOpen
  \bibfield  {author} {\bibinfo {author} {\bibfnamefont {T.}~\bibnamefont {Noda}}, \bibinfo {author} {\bibfnamefont {K.}~\bibnamefont {Kudo}}, \bibinfo {author} {\bibfnamefont {M.}~\bibnamefont {Takasuga}}, \bibinfo {author} {\bibfnamefont {M.}~\bibnamefont {Nohara}}, \bibinfo {author} {\bibfnamefont {T.}~\bibnamefont {Sugimoto}}, \bibinfo {author} {\bibfnamefont {D.}~\bibnamefont {Ootsuki}}, \bibinfo {author} {\bibfnamefont {M.}~\bibnamefont {Kobayashi}}, \bibinfo {author} {\bibfnamefont {K.}~\bibnamefont {Horiba}}, \bibinfo {author} {\bibfnamefont {K.}~\bibnamefont {Ono}}, \bibinfo {author} {\bibfnamefont {H.}~\bibnamefont {Kumigashira}}, \bibinfo {author} {\bibfnamefont {A.}~\bibnamefont {Fujimori}}, \bibinfo {author} {\bibfnamefont {N.~L.}\ \bibnamefont {Saini}},\ and\ \bibinfo {author} {\bibfnamefont {T.}~\bibnamefont {Mizokawa}},\ }\href {https://doi.org/10.7566/JPSJ.86.064708} {\bibfield  {journal} {\bibinfo  {journal} {Journal of the Physical Society of Japan}\ }\textbf {\bibinfo {volume} {86}},\
  \bibinfo {pages} {064708} (\bibinfo {year} {2017})}\BibitemShut {NoStop}%
\bibitem [{\citenamefont {Kreyssig}\ \emph {et~al.}(2008)\citenamefont {Kreyssig}, \citenamefont {Green}, \citenamefont {Lee}, \citenamefont {Samolyuk}, \citenamefont {Zajdel}, \citenamefont {Lynn}, \citenamefont {Bud\textquotesingle{ko}}, \citenamefont {Torikachvili}, \citenamefont {Ni}, \citenamefont {Nandi}, \citenamefont {Le\~ao}, \citenamefont {Poulton}, \citenamefont {Argyriou}, \citenamefont {Harmon}, \citenamefont {McQueeney}, \citenamefont {Canfield},\ and\ \citenamefont {Goldman}}]{Kreyssig08}%
  \BibitemOpen
  \bibfield  {author} {\bibinfo {author} {\bibfnamefont {A.}~\bibnamefont {Kreyssig}}, \bibinfo {author} {\bibfnamefont {M.~A.}\ \bibnamefont {Green}}, \bibinfo {author} {\bibfnamefont {Y.}~\bibnamefont {Lee}}, \bibinfo {author} {\bibfnamefont {G.~D.}\ \bibnamefont {Samolyuk}}, \bibinfo {author} {\bibfnamefont {P.}~\bibnamefont {Zajdel}}, \bibinfo {author} {\bibfnamefont {J.~W.}\ \bibnamefont {Lynn}}, \bibinfo {author} {\bibfnamefont {S.~L.}\ \bibnamefont {Bud\textquotesingle{ko}}}, \bibinfo {author} {\bibfnamefont {M.~S.}\ \bibnamefont {Torikachvili}}, \bibinfo {author} {\bibfnamefont {N.}~\bibnamefont {Ni}}, \bibinfo {author} {\bibfnamefont {S.}~\bibnamefont {Nandi}}, \bibinfo {author} {\bibfnamefont {J.~B.}\ \bibnamefont {Le\~ao}}, \bibinfo {author} {\bibfnamefont {S.~J.}\ \bibnamefont {Poulton}}, \bibinfo {author} {\bibfnamefont {D.~N.}\ \bibnamefont {Argyriou}}, \bibinfo {author} {\bibfnamefont {B.~N.}\ \bibnamefont {Harmon}}, \bibinfo {author} {\bibfnamefont {R.~J.}\ \bibnamefont {McQueeney}}, \bibinfo
  {author} {\bibfnamefont {P.~C.}\ \bibnamefont {Canfield}},\ and\ \bibinfo {author} {\bibfnamefont {A.~I.}\ \bibnamefont {Goldman}},\ }\href {https://doi.org/10.1103/PhysRevB.78.184517} {\bibfield  {journal} {\bibinfo  {journal} {Physical Review B}\ }\textbf {\bibinfo {volume} {78}},\ \bibinfo {pages} {184517} (\bibinfo {year} {2008})}\BibitemShut {NoStop}%
\bibitem [{\citenamefont {Mewis}\ and\ \citenamefont {Distler}(1980)}]{Mewis80}%
  \BibitemOpen
  \bibfield  {author} {\bibinfo {author} {\bibfnamefont {A.}~\bibnamefont {Mewis}}\ and\ \bibinfo {author} {\bibfnamefont {A.}~\bibnamefont {Distler}},\ }\href {https://doi.org/doi:10.1515/znb-1980-0326} {\bibfield  {journal} {\bibinfo  {journal} {Zeitschrift f\"{u}r Naturforschung B}\ }\textbf {\bibinfo {volume} {35}},\ \bibinfo {pages} {391} (\bibinfo {year} {1980})}\BibitemShut {NoStop}%
\bibitem [{Cry(2015)}]{CrysalysPro}%
  \BibitemOpen
  \href@noop {} {\bibinfo {title} {{Rigaku Oxford Diffraction Ltd, Yarnton, Oxfordshire, E 2015 CrysAlis PRO}}} (\bibinfo {year} {2015})\BibitemShut {NoStop}%
\bibitem [{\citenamefont {Sheldrick}(2008)}]{Sheldrick08}%
  \BibitemOpen
  \bibfield  {author} {\bibinfo {author} {\bibfnamefont {G.~M.}\ \bibnamefont {Sheldrick}},\ }\href@noop {} {\bibfield  {journal} {\bibinfo  {journal} {Acta Crystallographica Section A: Foundations of Crystallography}\ }\textbf {\bibinfo {volume} {64}},\ \bibinfo {pages} {112} (\bibinfo {year} {2008})}\BibitemShut {NoStop}%
\bibitem [{\citenamefont {Pet\v{r}\'{i}\v{c}ek}\ \emph {et~al.}(2014)\citenamefont {Pet\v{r}\'{i}\v{c}ek}, \citenamefont {Du\v{s}ek},\ and\ \citenamefont {Palatinus}}]{Petříček14}%
  \BibitemOpen
  \bibfield  {author} {\bibinfo {author} {\bibfnamefont {V.}~\bibnamefont {Pet\v{r}\'{i}\v{c}ek}}, \bibinfo {author} {\bibfnamefont {M.}~\bibnamefont {Du\v{s}ek}},\ and\ \bibinfo {author} {\bibfnamefont {L.}~\bibnamefont {Palatinus}},\ }\href {https://doi.org/doi:10.1515/zkri-2014-1737} {\bibfield  {journal} {\bibinfo  {journal} {Zeitschrift f\"{u}r Kristallographie - Crystalline Materials}\ }\textbf {\bibinfo {volume} {229}},\ \bibinfo {pages} {345} (\bibinfo {year} {2014})}\BibitemShut {NoStop}%
\bibitem [{\citenamefont {Girard}\ \emph {et~al.}(2019)\citenamefont {Girard}, \citenamefont {Nguyen-Thanh}, \citenamefont {Souliou}, \citenamefont {Stekiel}, \citenamefont {Morgenroth}, \citenamefont {Paolasini}, \citenamefont {Minelli}, \citenamefont {Gambetti}, \citenamefont {Winkler},\ and\ \citenamefont {Bosak}}]{Girard19}%
  \BibitemOpen
  \bibfield  {author} {\bibinfo {author} {\bibfnamefont {A.}~\bibnamefont {Girard}}, \bibinfo {author} {\bibfnamefont {T.}~\bibnamefont {Nguyen-Thanh}}, \bibinfo {author} {\bibfnamefont {S.}~\bibnamefont {Souliou}}, \bibinfo {author} {\bibfnamefont {M.}~\bibnamefont {Stekiel}}, \bibinfo {author} {\bibfnamefont {W.}~\bibnamefont {Morgenroth}}, \bibinfo {author} {\bibfnamefont {L.}~\bibnamefont {Paolasini}}, \bibinfo {author} {\bibfnamefont {A.}~\bibnamefont {Minelli}}, \bibinfo {author} {\bibfnamefont {D.}~\bibnamefont {Gambetti}}, \bibinfo {author} {\bibfnamefont {B.}~\bibnamefont {Winkler}},\ and\ \bibinfo {author} {\bibfnamefont {A.}~\bibnamefont {Bosak}},\ }\href@noop {} {\bibfield  {journal} {\bibinfo  {journal} {Journal of Synchrotron Radiation}\ }\textbf {\bibinfo {volume} {26}},\ \bibinfo {pages} {272} (\bibinfo {year} {2019})}\BibitemShut {NoStop}%
\bibitem [{\citenamefont {Sefat}\ \emph {et~al.}(2009)\citenamefont {Sefat}, \citenamefont {McGuire}, \citenamefont {Jin}, \citenamefont {Sales}, \citenamefont {Mandrus}, \citenamefont {Ronning}, \citenamefont {Bauer},\ and\ \citenamefont {Mozharivskyj}}]{Sefat09}%
  \BibitemOpen
  \bibfield  {author} {\bibinfo {author} {\bibfnamefont {A.~S.}\ \bibnamefont {Sefat}}, \bibinfo {author} {\bibfnamefont {M.~A.}\ \bibnamefont {McGuire}}, \bibinfo {author} {\bibfnamefont {R.}~\bibnamefont {Jin}}, \bibinfo {author} {\bibfnamefont {B.~C.}\ \bibnamefont {Sales}}, \bibinfo {author} {\bibfnamefont {D.}~\bibnamefont {Mandrus}}, \bibinfo {author} {\bibfnamefont {F.}~\bibnamefont {Ronning}}, \bibinfo {author} {\bibfnamefont {E.~D.}\ \bibnamefont {Bauer}},\ and\ \bibinfo {author} {\bibfnamefont {Y.}~\bibnamefont {Mozharivskyj}},\ }\href {https://doi.org/10.1103/PhysRevB.79.094508} {\bibfield  {journal} {\bibinfo  {journal} {Phys. Rev. B}\ }\textbf {\bibinfo {volume} {79}},\ \bibinfo {pages} {094508} (\bibinfo {year} {2009})}\BibitemShut {NoStop}%
\bibitem [{\citenamefont {Dulong}\ and\ \citenamefont {Petit}(1819)}]{DulongPetit1819}%
  \BibitemOpen
  \bibfield  {author} {\bibinfo {author} {\bibfnamefont {P.~L.}\ \bibnamefont {Dulong}}\ and\ \bibinfo {author} {\bibfnamefont {A.-T.}\ \bibnamefont {Petit}},\ }\href@noop {} {\emph {\bibinfo {title} {Recherches sur quelques points importans de la theorie de la chaleur}}}\ (\bibinfo {year} {1819})\BibitemShut {NoStop}%
\bibitem [{\citenamefont {Willa}\ \emph {et~al.}(2018)\citenamefont {Willa}, \citenamefont {Willa}, \citenamefont {Song}, \citenamefont {Gu}, \citenamefont {Schneeloch}, \citenamefont {Zhong}, \citenamefont {Koshelev}, \citenamefont {Kwok},\ and\ \citenamefont {Welp}}]{Willa18}%
  \BibitemOpen
  \bibfield  {author} {\bibinfo {author} {\bibfnamefont {K.}~\bibnamefont {Willa}}, \bibinfo {author} {\bibfnamefont {R.}~\bibnamefont {Willa}}, \bibinfo {author} {\bibfnamefont {K.~W.}\ \bibnamefont {Song}}, \bibinfo {author} {\bibfnamefont {G.~D.}\ \bibnamefont {Gu}}, \bibinfo {author} {\bibfnamefont {J.~A.}\ \bibnamefont {Schneeloch}}, \bibinfo {author} {\bibfnamefont {R.}~\bibnamefont {Zhong}}, \bibinfo {author} {\bibfnamefont {A.~E.}\ \bibnamefont {Koshelev}}, \bibinfo {author} {\bibfnamefont {W.-K.}\ \bibnamefont {Kwok}},\ and\ \bibinfo {author} {\bibfnamefont {U.}~\bibnamefont {Welp}},\ }\href {https://doi.org/10.1103/PhysRevB.98.184509} {\bibfield  {journal} {\bibinfo  {journal} {Physical Review B}\ }\textbf {\bibinfo {volume} {98}},\ \bibinfo {pages} {184509} (\bibinfo {year} {2018})}\BibitemShut {NoStop}%
\bibitem [{\citenamefont {Bardeen}\ \emph {et~al.}(1957)\citenamefont {Bardeen}, \citenamefont {Cooper},\ and\ \citenamefont {Schrieffer}}]{BCS57}%
  \BibitemOpen
  \bibfield  {author} {\bibinfo {author} {\bibfnamefont {J.}~\bibnamefont {Bardeen}}, \bibinfo {author} {\bibfnamefont {L.~N.}\ \bibnamefont {Cooper}},\ and\ \bibinfo {author} {\bibfnamefont {J.~R.}\ \bibnamefont {Schrieffer}},\ }\href {https://doi.org/10.1103/PhysRev.108.1175} {\bibfield  {journal} {\bibinfo  {journal} {Physical Review}\ }\textbf {\bibinfo {volume} {108}},\ \bibinfo {pages} {1175} (\bibinfo {year} {1957})}\BibitemShut {NoStop}%
\bibitem [{\citenamefont {Kudo}\ \emph {et~al.}(2017)\citenamefont {Kudo}, \citenamefont {Takasuga},\ and\ \citenamefont {Nohara}}]{Kudo17}%
  \BibitemOpen
  \bibfield  {author} {\bibinfo {author} {\bibfnamefont {K.}~\bibnamefont {Kudo}}, \bibinfo {author} {\bibfnamefont {M.}~\bibnamefont {Takasuga}},\ and\ \bibinfo {author} {\bibfnamefont {M.}~\bibnamefont {Nohara}},\ }\href {https://arxiv.org/abs/1704.04854} {} (\bibinfo {year} {2017}),\ \Eprint {https://arxiv.org/abs/1704.04854} {arXiv:1704.04854 [cond-mat.supr-con]} \BibitemShut {NoStop}%
\bibitem [{\citenamefont {Eckberg}\ \emph {et~al.}(2018)\citenamefont {Eckberg}, \citenamefont {Wang}, \citenamefont {Hodovanets}, \citenamefont {Kim}, \citenamefont {Campbell}, \citenamefont {Zavalij}, \citenamefont {Piccoli},\ and\ \citenamefont {Paglione}}]{Eckberg18}%
  \BibitemOpen
  \bibfield  {author} {\bibinfo {author} {\bibfnamefont {C.}~\bibnamefont {Eckberg}}, \bibinfo {author} {\bibfnamefont {L.}~\bibnamefont {Wang}}, \bibinfo {author} {\bibfnamefont {H.}~\bibnamefont {Hodovanets}}, \bibinfo {author} {\bibfnamefont {H.}~\bibnamefont {Kim}}, \bibinfo {author} {\bibfnamefont {D.~J.}\ \bibnamefont {Campbell}}, \bibinfo {author} {\bibfnamefont {P.}~\bibnamefont {Zavalij}}, \bibinfo {author} {\bibfnamefont {P.}~\bibnamefont {Piccoli}},\ and\ \bibinfo {author} {\bibfnamefont {J.}~\bibnamefont {Paglione}},\ }\href {https://doi.org/10.1103/PhysRevB.97.224505} {\bibfield  {journal} {\bibinfo  {journal} {Phys. Rev. B}\ }\textbf {\bibinfo {volume} {97}},\ \bibinfo {pages} {224505} (\bibinfo {year} {2018})}\BibitemShut {NoStop}%
\bibitem [{\citenamefont {Hardy}\ \emph {et~al.}(2020)\citenamefont {Hardy}, \citenamefont {Doussoulin}, \citenamefont {Klein}, \citenamefont {He}, \citenamefont {Demuer}, \citenamefont {Willa}, \citenamefont {Willa}, \citenamefont {Haghighirad}, \citenamefont {Wolf}, \citenamefont {Merz}, \citenamefont {Meingast},\ and\ \citenamefont {Marcenat}}]{Hardy20}%
  \BibitemOpen
  \bibfield  {author} {\bibinfo {author} {\bibfnamefont {F.}~\bibnamefont {Hardy}}, \bibinfo {author} {\bibfnamefont {L.}~\bibnamefont {Doussoulin}}, \bibinfo {author} {\bibfnamefont {T.}~\bibnamefont {Klein}}, \bibinfo {author} {\bibfnamefont {M.}~\bibnamefont {He}}, \bibinfo {author} {\bibfnamefont {A.}~\bibnamefont {Demuer}}, \bibinfo {author} {\bibfnamefont {R.}~\bibnamefont {Willa}}, \bibinfo {author} {\bibfnamefont {K.}~\bibnamefont {Willa}}, \bibinfo {author} {\bibfnamefont {A.~A.}\ \bibnamefont {Haghighirad}}, \bibinfo {author} {\bibfnamefont {T.}~\bibnamefont {Wolf}}, \bibinfo {author} {\bibfnamefont {M.}~\bibnamefont {Merz}}, \bibinfo {author} {\bibfnamefont {C.}~\bibnamefont {Meingast}},\ and\ \bibinfo {author} {\bibfnamefont {C.}~\bibnamefont {Marcenat}},\ }\href {https://doi.org/10.1103/PhysRevResearch.2.033319} {\bibfield  {journal} {\bibinfo  {journal} {Physical Review Research}\ }\textbf {\bibinfo {volume} {2}},\ \bibinfo {pages} {033319} (\bibinfo {year} {2020})}\BibitemShut {NoStop}%
\bibitem [{\citenamefont {Willa}\ \emph {et~al.}(2019)\citenamefont {Willa}, \citenamefont {Willa}, \citenamefont {Bao}, \citenamefont {Koshelev}, \citenamefont {Chung}, \citenamefont {Kanatzidis}, \citenamefont {Kwok},\ and\ \citenamefont {Welp}}]{Willa19}%
  \BibitemOpen
  \bibfield  {author} {\bibinfo {author} {\bibfnamefont {K.}~\bibnamefont {Willa}}, \bibinfo {author} {\bibfnamefont {R.}~\bibnamefont {Willa}}, \bibinfo {author} {\bibfnamefont {J.~K.}\ \bibnamefont {Bao}}, \bibinfo {author} {\bibfnamefont {A.~E.}\ \bibnamefont {Koshelev}}, \bibinfo {author} {\bibfnamefont {D.~Y.}\ \bibnamefont {Chung}}, \bibinfo {author} {\bibfnamefont {M.~G.}\ \bibnamefont {Kanatzidis}}, \bibinfo {author} {\bibfnamefont {W.~K.}\ \bibnamefont {Kwok}},\ and\ \bibinfo {author} {\bibfnamefont {U.}~\bibnamefont {Welp}},\ }\href {https://doi.org/10.1103/PhysRevB.99.180502} {\bibfield  {journal} {\bibinfo  {journal} {Physical Review B}\ }\textbf {\bibinfo {volume} {99}},\ \bibinfo {pages} {180502} (\bibinfo {year} {2019})}\BibitemShut {NoStop}%
\bibitem [{\citenamefont {Prozorov}\ \emph {et~al.}(2024)\citenamefont {Prozorov}, \citenamefont {Kogan}, \citenamefont {Ko\'{n}czykowski},\ and\ \citenamefont {Tanatar}}]{Prozorov24}%
  \BibitemOpen
  \bibfield  {author} {\bibinfo {author} {\bibfnamefont {R.}~\bibnamefont {Prozorov}}, \bibinfo {author} {\bibfnamefont {V.~G.}\ \bibnamefont {Kogan}}, \bibinfo {author} {\bibfnamefont {M.}~\bibnamefont {Ko\'{n}czykowski}},\ and\ \bibinfo {author} {\bibfnamefont {M.~A.}\ \bibnamefont {Tanatar}},\ }\href {https://doi.org/10.1103/PhysRevB.109.024506} {\bibfield  {journal} {\bibinfo  {journal} {Physical Review B}\ }\textbf {\bibinfo {volume} {109}},\ \bibinfo {pages} {024506} (\bibinfo {year} {2024})}\BibitemShut {NoStop}%
\bibitem [{\citenamefont {Werthamer}\ \emph {et~al.}(1966)\citenamefont {Werthamer}, \citenamefont {Helfand},\ and\ \citenamefont {Hohenberg}}]{Werthamer66}%
  \BibitemOpen
  \bibfield  {author} {\bibinfo {author} {\bibfnamefont {N.~R.}\ \bibnamefont {Werthamer}}, \bibinfo {author} {\bibfnamefont {E.}~\bibnamefont {Helfand}},\ and\ \bibinfo {author} {\bibfnamefont {P.~C.}\ \bibnamefont {Hohenberg}},\ }\href {https://doi.org/10.1103/PhysRev.147.295} {\bibfield  {journal} {\bibinfo  {journal} {Physical Review}\ }\textbf {\bibinfo {volume} {147}},\ \bibinfo {pages} {295} (\bibinfo {year} {1966})}\BibitemShut {NoStop}%
\bibitem [{\citenamefont {Park}\ \emph {et~al.}(2010)\citenamefont {Park}, \citenamefont {Lee}, \citenamefont {Bauer}, \citenamefont {Thompson},\ and\ \citenamefont {Ronning}}]{Park_2010}%
  \BibitemOpen
  \bibfield  {author} {\bibinfo {author} {\bibfnamefont {T.}~\bibnamefont {Park}}, \bibinfo {author} {\bibfnamefont {H.}~\bibnamefont {Lee}}, \bibinfo {author} {\bibfnamefont {E.~D.}\ \bibnamefont {Bauer}}, \bibinfo {author} {\bibfnamefont {J.~D.}\ \bibnamefont {Thompson}},\ and\ \bibinfo {author} {\bibfnamefont {F.}~\bibnamefont {Ronning}},\ }\href {https://doi.org/10.1088/1742-6596/200/1/012155} {\bibfield  {journal} {\bibinfo  {journal} {Journal of Physics: Conference Series}\ }\textbf {\bibinfo {volume} {200}},\ \bibinfo {pages} {012155} (\bibinfo {year} {2010})}\BibitemShut {NoStop}%
\bibitem [{\citenamefont {Kito}\ \emph {et~al.}(2008)\citenamefont {Kito}, \citenamefont {Eisaki},\ and\ \citenamefont {Iyo}}]{Kito08}%
  \BibitemOpen
  \bibfield  {author} {\bibinfo {author} {\bibfnamefont {H.}~\bibnamefont {Kito}}, \bibinfo {author} {\bibfnamefont {H.}~\bibnamefont {Eisaki}},\ and\ \bibinfo {author} {\bibfnamefont {A.}~\bibnamefont {Iyo}},\ }\href {https://doi.org/10.1143/JPSJ.77.063707} {\bibfield  {journal} {\bibinfo  {journal} {Journal of the Physical Society of Japan}\ }\textbf {\bibinfo {volume} {77}},\ \bibinfo {pages} {063707} (\bibinfo {year} {2008})}\BibitemShut {NoStop}%
\bibitem [{\citenamefont {Tomioka}\ \emph {et~al.}(2008)\citenamefont {Tomioka}, \citenamefont {Ito}, \citenamefont {Kito}, \citenamefont {Iyo}, \citenamefont {Eisaki}, \citenamefont {Ishida}, \citenamefont {Nakajima},\ and\ \citenamefont {Uchida}}]{Tomioka08}%
  \BibitemOpen
  \bibfield  {author} {\bibinfo {author} {\bibfnamefont {Y.}~\bibnamefont {Tomioka}}, \bibinfo {author} {\bibfnamefont {T.}~\bibnamefont {Ito}}, \bibinfo {author} {\bibfnamefont {H.}~\bibnamefont {Kito}}, \bibinfo {author} {\bibfnamefont {A.}~\bibnamefont {Iyo}}, \bibinfo {author} {\bibfnamefont {H.}~\bibnamefont {Eisaki}}, \bibinfo {author} {\bibfnamefont {S.}~\bibnamefont {Ishida}}, \bibinfo {author} {\bibfnamefont {M.}~\bibnamefont {Nakajima}},\ and\ \bibinfo {author} {\bibfnamefont {S.-i.}\ \bibnamefont {Uchida}},\ }\href {https://doi.org/10.1143/JPSJS.77SC.136} {\bibfield  {journal} {\bibinfo  {journal} {Journal of the Physical Society of Japan}\ }\textbf {\bibinfo {volume} {77}},\ \bibinfo {pages} {136} (\bibinfo {year} {2008})}\BibitemShut {NoStop}%
\end{thebibliography}%
\end{document}